\RequirePackage{fix-cm}
\documentclass{svjour3}                     % onecolumn (standard format)
\smartqed  % flush right qed marks, e.g. at end of proof
\usepackage[utf8x]{inputenc}			% Zeichenausgabe (vor fontenc!)
\usepackage[T1]{fontenc}				% Zeichenkodierung
\usepackage[english]{babel}				% Sprachunterstützung
\usepackage{microtype}					% Verbesserter Satzspiegel
\usepackage{amsmath}					% erweiterte Mathe-Umgebung
\usepackage{blindtext}					% Blindtext
\usepackage{booktabs}					% Tabellenlayout
\usepackage{color}						% Farbdef. für farbige Links
\usepackage{nomencl}					% Nomenklatur
\usepackage{caption}					% Bildunterschriften
\usepackage{subcaption}					% Teilunterschriften
\usepackage{bm}	
\usepackage{colortbl}
\usepackage{amssymb}
\usepackage{here} 
\definecolor{dunkelgrau}{rgb}{0.8,0.8,0.8}
\definecolor{hellgrau}{rgb}{0.95,0.95,0.95}
\usepackage{multirow}

\usepackage{amsfonts}

\usepackage[numbers,sort&compress]{natbib}

\usepackage{lmodern}
\usepackage{amsmath}

\usepackage[colorlinks=true,citecolor=red,linkcolor=black]{hyperref}

\usepackage[T1]{fontenc}
\usepackage{xcolor}
\usepackage{lmodern}
\usepackage{listings}

\usepackage{graphicx}
\usepackage{hyperref}

\usepackage{lineno}

\begin{document}

	%\linenumbers
	
\title{Improved Perception of AEC Construction Details via Immersive Teaching in Virtual Reality}
\titlerunning{Teaching AEC Construction Details via VR}        
	
\author{M.\,A. Kraus* \and R. Rust \and M. Rietschel \and D. Hall}
	
\institute{
    *corresponding author: Dr. Michael A. Kraus \\
    Dr. Michael A. Kraus, M.Sc.(hons) and Dr. Romana Rust \at  ETH Zurich\\ %
	Design++ Initiative and Immersive Design Lab, ETH~Zurich\\
	Stefano-Franscini-Platz 5\\
	8093 Zürich \\
	\email{kraus@ibk.baug.ethz.ch}\\
\\
	M.Sc. Maximilian Rietschel and Prof. Dr. Daniel Hall \at  ETH Zurich \\ %
	Innovative \& Industrial Construction\\
}

\date{Preprint - Version 21.09.2022. Submitted for Review.}

\maketitle

\begin{abstract}
	This work proposes, implements and tests an immersive framework upon Virtual Reality (VR) for comprehension, knowledge development and learning process assisting an improved perception of complex spatial arrangements in architecture and civil engineering in comparison to the traditional 2D projection drawing-based method. The research focuses on the prototypical example of construction details as a traditionally difficult teaching task for conveying geometric and semantic information to students. Our mixed-methods study analyses test results of two test panel groups (test group used VR, control group used 2D drawings) upon different questions about geometric and functional aspects of the construction detail as well as surveys and interviews of participating lecturers, students and laypersons towards their experience using the VR tool. The quantitative analysis of the test results prove that for participants with reasonable prior knowledge, there is no significant difference between the test and control groups towards the test results. However, for participants with little pre-existing knowledge (such as novice architecture and civil engineering students), a significantly better learning score for the test group is detected. Moreover, both groups rated the VR experience as an enjoyable and engaging way of learning. Analysis of survey results towards the VR experience reveals, that students, lecturers and professionals alike enjoyed the VR experience more than traditional learning of the construction detail. During the post-experiment qualitative evaluation in the form of interviews, the panel expressed an improved understanding, increased enthusiasm for the topic, and greater desire for other topics to be presented using VR tools. The expressed better understanding of design concepts after the VR experience by the students is statistically significant on average in the exam results. The results of this study supports the authors’ core assumption, that the presentation of contextual 3D models is a promising teaching approach to illustrate content. To that end, VR technology will augment traditional teaching formats in architecture and civil engineering curricula in the near future.
\end{abstract}

\section{Introduction} \label{sec:intro}
Over the past years, Extended Reality (XR) and especially Virtual Reality (VR) technologies were becoming more affordable to a big consumer market. The dissemination of these technologies hence increases steadily and promises new opportunities and use cases. The majority of higher education institutions in the architecture, engineering and construction (AEC) sector however have not yet adopted new digital learning technologies and XR in particular, or have done so only to a relatively small extent. 

Students must possess sophisticated analytical thinking and abstraction skills in order to comprehend the material in introductory and foundational courses on the design and analysis of structures and the built environment as well as the profession's guiding principles. This abstraction in first year lectures is a barrier, particularly for students of architecture and civil engineering lacking relevant professional experience. Building detailing is a prototypical example area where students are taught the fundamentals of functional, reliable and efficient construction details. Today, the subject contents are delivered through instructor-led lectures, exercises, and textbooks based on two-dimensional drawings of three-dimensional building parts as sections and views. The content to be learned consists of considerations towards geometry, adjacency, functionality and construction sequences as construction details are usually fabricated in a complex joined fashion. Students are requested to develop a mental picture of the 3D structure together with its assembly of multiple members upon minimal information and expertise when projecting to 2D. In addition, current civil engineering curricula lack of lectures or exercises on drawing 2D representations of 3D phenomena in addition to a little number of construction site visits due to time, expense, and availability restrictions. However, both means would improve students' capacity to perceive three-dimensional groupings in two dimensions by complementing lecture material and providing genuine 3D content of construction details. In conclusion, the current method of teaching building details in AEC causes students to rather memorise the 2D representation rather than developing insight and knowledge of the structural detail, its functionality, and composition sequence. 

Motivated by our previous work \cite{kraus.2021struct,kraus.2022mixed} on using Augmented and Mixed Reality applications for developing novel workflows in teaching structural engineering methods, this paper investigates a workflow for teaching construction details with VR and measures its impact via a mixed-methods approach. Our past studies delivered results, which coincide with findings in literature such as \cite{Pan.2006,fogarty2018improving,veurink2011raising}. These studies prove, that with improving spatial visualisation and virtual learning environments, students' understanding can be enhanced, motivated and stimulated for science, technology, engineering, and math (STEM) subjects in general as well as for construction and analysis of AEC problems specifically. 

The objective of this mixed-methods research is twofold: (i) to develop a technical workflow for embedding construction detail analysis content into an immersive VR application, (ii) to conduct a test and survey amongst the study panel (consisting of lecturers, students and laypersons) to measure the acceptance, ergonomics and learning impact of the developed VR application for a prototypical construction detail. The VR framework is intended to significantly improve comprehension, knowledge development and the learning process of complex spatial arrangements in architecture and civil engineering in comparison to the traditional 2D projection drawing-based method. The VR models provide interactive and exciting means of interaction for layperson as well as professionals, where the main focus of this study lays on improving students' understanding of complex spatial arrangements together with the semantics of the different building elements. The research focuses on the prototypical example of construction details as a traditionally difficult teaching task for conveying geometric and semantic information to students. This study assessed and evaluated the fitness of the VR tool to provide a meaningful, supportive and efficient immersive lecturing method for construction details. Besides the quantification of knowledge gain, this study evaluated the perception of academics (students and faculty) as well as professionals (architects and civil engineers in practice) towards the impact and implications of using such a VR tool over traditional lecturing methods to convey complex built structures. Furthermore, the qualitative analysis investigates the effect of interactivity, i.e. users' ability to zoom in, view animations from alternative perspectives, and stop, start or control the speed of animation, as well as the effect of immersion using a VR headset. In total, three research questions are considered: (Q1): Does VR improve student's understanding of complex spatial construction arrangements in contrast to 2D drawings?; (Q2): How does prior experience affect the VR learning gains?; and (Q3): How is the VR tool perceived by students, instructors but also AEC professionals towards ergonomy and usefulness as well as its future potential in education. The test panel is split into a test group (TG-VR) and a control group (CG-2D) during the experiment, whereas the final survey is provided to the whole panel. 

In the remainder of this paper, at first a review of relevant literature is presented in Sec.~\ref{Background} and Sec.~\ref{sec:Methods} summaries the main methods. In Sec.~\ref{sec:Limitations} we highlight the limitations, Sec.~\ref{sec:Results} provides the results of our implementation as well as the conducted experiments amongst test and control group. In Sec.~\ref{sec:Conclusion} we provide a discussion of the results next to an interpretation with conclusions. An outlook on future activities and open research points concludes this paper. Additional material is provided in the appendix Sec.~\ref{sec:Appendix} and supplemental material sections Sec.~\ref{sec:Supplemental}.

\section{State-of-the-Art and Background}  \label{Background}
The term extended reality (XR) comprises all technologies for combining real and virtual environments as well as human-machine interactions generated by computer technology and wearables \cite{fast2018testing,OsortoCarrasco2021}. XR technologies create immersive digital worlds: 
\begin{itemize}
	\item Augmented Reality (AR): the real world is enhanced with digital content
	\item Virtual Reality (VR): the user is immersed into a fully digital environment fading the real environment completely
	\item Mixed Reality (MR): computer generated content is blended in varying proportion with an individual's view of the real-world scene with additional real-time interaction
\end{itemize}

\begin{figure} [h]
	\centering
	\includegraphics[width=0.98\linewidth]{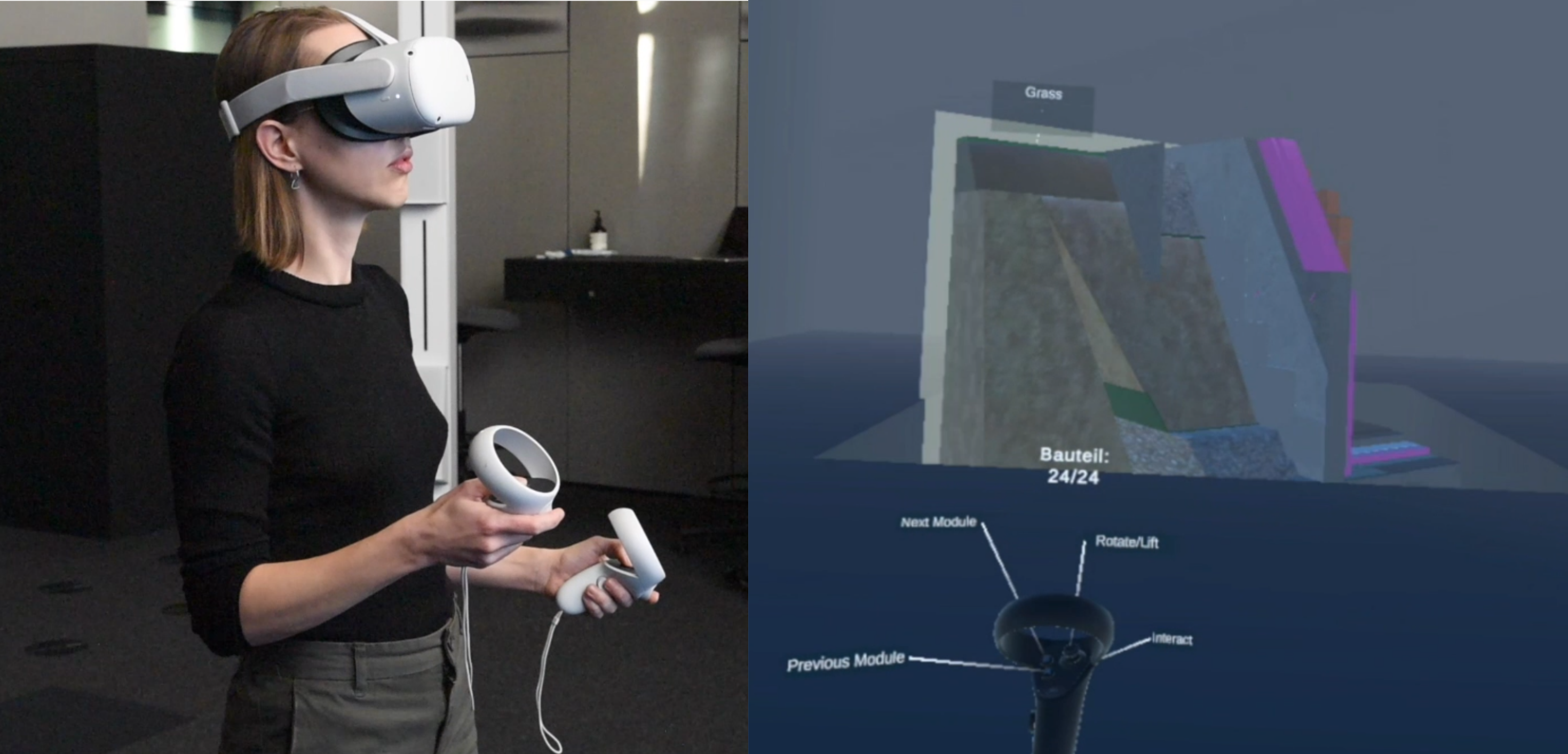} \\
	\caption{User (left) in the immersive VR environment (right) used in this study (image courtesy of Maximilian Rietschel)}
	\label{fig:Concept}
\end{figure}

The continuing decline in prices with increased availability for most XR tools allowed individuals to start exploring possibilities of XR devices in different fields for personal or professional purposes. VR is a powerful tool for comprehending and analysing complicated three-dimensional setups by providing an immersive environment allowing for users interaction through spacial interaction such as walking around a model to gain overview and experience the size and scope of significant aspects in a natural and tangible manner, cf. Fig.~\ref{fig:Concept}. In addition, interactive content may further be provided to trigger user interaction or providing more information. \cite{kraus.2021struct} provides an overview on numerous studies towards applications of XR in AEC practice, indicated the promising potential of XR in AEC. However, there is only limited research on the development and integration of XR technologies into undergraduate AEC teaching, including the incorporation of gamification and cyber teaching tools \cite{dinis2017virtual,kraus.2021struct}.

According to previous research \cite{felder2004a,felder2004b}, traditional teacher-based lecturing is not the most effective method of instruction since it fails to intrinsically motivate students and offers few, if any, opportunities for them to add to their prior knowledge. Especially students in architecture and civil engineering are interested in gaining knowledge on how to design, verify and construct different structures of the built environment. Due to the limitations of conventional teaching methods, particularly the disconnect of classroom learning and real-world applications in addition to a lack of hands-on opportunities, current AEC education largely fails to provide students the opportunity to experience and understand the profession on a larger, application-based scale \cite{Chen2014}. Today, instruction focuses on small-scale problems, which can be handled by analytical approaches, as opposed to a comprehensive resp. holistic perspective of structures or components in a collaborative setting \cite{ferguson1993engineering}. 

Communicating technical matters and designs in current AEC practise is vastly based on 2D projection drawings employing sectional views either on paper or in digital from, despite that design modelling nowadays is mostly conducted in 3D. To that end, a 2D drawing of a structural configuration is an abstraction of the 3D object and associated attributes, while the projection causes the loss of some information. This abstraction is especially hard for AEC students due to inexperience and unfamiliarity with the current approach. Despite the value of learning to abstract 2D drawings from 3D building objects for students, \cite{sorby2009developing} provides evidence for the special significance of 3D visualisation capabilities for architects and civil engineers, which can be greatly supported by XR methods. Acc. to \cite{shirazi2015a,Sampaio2014} XR facilitates learning of abstract and difficult-to-understand topics by reducing or even eliminating ambiguities by using 3D visualisations as well as real-time interactions. The influence of studying objects in either a 2D or 3D environment towards the spatial understanding was looked at by \cite{schnabel2003spatial}. The study measured the accuracy of the reconstructed objects during interviews for two different groups. The outcome indicated immersive 3D experiences to provide users with improved comprehension of complex volumes together with their spatial arrangements. \cite{fogarty2018improving} investigates the use of VR tools to aid student comprehension of complex spatial arrangements in civil engineering with a focus on structural buckling responses. Students were provided VR models for different structures and buckling modes. The study reported that students are able to identify and visualise buckling modes more accurately after the VR experience. Both, instructors and students note the advantages of using VR for explaining and understanding the complex deformation modes of buckling structural members, which manifests in an - on average - statistically significant improvement on the post-test evaluation of students' comprehension.

\cite{Messner.2003} created and evaluated a VR interface for developing construction plans of a nuclear power plant in less than an hour in AEC undergraduate programs. The study demonstrated the value of immersive VR displays for this kind of presentation, and the technology enabled students to comprehend planning challenges beyond their prior knowledge and visualising abilities of structures and infrastructure. An educational XR application is provided by \cite{liarokapis.2004a} to allow user interaction with 3D material using web technology and AR/VR approaches. \cite{Maghool.2018} developed a VR application to provide architecture students the opportunity to experience being on a construction site with the opportunity to closely examine details or evaluate their knowledge. The study revealed, that the current AEC teaching practice leaves out a significant portion of learner types. Using VR was suggested to support a problem-based and experiential learning to overcome this situation in contrast to traditional teaching techniques. \cite{Elgewely:2021} also observe a lack of experiential learning in architectural education today, which is mainly attributed to the low number of site visits despite their importance of being a valuable extension of classroom activity. An educational VR environment (VRE) with BIM technology integration is developed and assessed in order to improve the situation and to tackle the critique of \cite{Maghool.2018} towards the time consuming and tedious nature of integration of building information into VRE. \cite{kraus.2021struct} and \cite{kraus.2022mixed} take a different approach: instead of simulating a construction site, they developed a new approach to teaching in the classroom, where the lecture is complemented by an AR application, where students can take an in-depth look at 3D structural engineering details with supplemental information. The study empirically found, that XR technologies possess great potential for improving effectiveness of teaching by displaying environments and associated information in an intuitive way similar to real objects. The study also observes, that students and instructors reported higher enjoyment of the learning process when using the XR technology.

In addition to better learning by the students, VR also offers benefits in terms of time and cost savings. In addition to the findings of \cite{Maghool.2018} towards learner types, \cite{Gutierrez2015} point out the opportunities of VR towards allowing student to learn according to their own pace and needs. \cite{Hafner.2013} created a university-level course to educate students how to use VR hardware, software, and applications in engineering. The study discovered a higher motivation amongst students on completing assigned tasks when VR was employed.\cite{dinis2017virtual} developed VR and AR applications for students of an introductory class of the Integrated Masters in Civil Engineering and tested those in two trials. Further successful development of VR applications in design and education tasks in AEC are reported by \cite{Sampaio2014,Wolfartsberger2019}. The influence of the order of intervention, i.e. the exposure to traditional teaching methods versus the mixed-reality method, as well as the role of embodiment, i.e. body motions, gestures, and the sense of immersion in the XR environment, onto students' comprehension was investigated by \cite{johnson2014collaborative} as well as \cite{kraus.2021struct} by a mixed-methods approach combining qualitative and quantitative evaluations. Both studies provide evidence, that exposure to an embodied XR experience results in greater student learning gains, although varying degrees of embodiment were not studied for the difference in learning gains. The changes in the roles of lecturers and students induced by XR methods and simulation games is discussed in \cite{Deshpande2011}. In this study, instructors act as promoters of knowledge, and students receive a more active and collaborative role. This is in agreement with the conclusions in \cite{kraus.2021struct}, where XR methods are reported to enable students to not only work interactively in groups and under supervision of the lecturer, but also to use the tool individually at home to review and reinforce the lecture materials. 

We found little literature for the use of tools to the VR models described in this work for architecture and civil engineering education. Hence, the development and thorough analysis provides a unique chance for dissemination into related fields such as mechanical engineering.

% Based on the literature review and the obtained conclusions, this research selects appropriate methods and proposes to develop, implement and test an immersive framework upon VR for knowledge development and learning process assisting for the prototypical example of construction details in order to assess acceptance, ergonomics and learning impact of VR compared to traditional 2D drawing-based methods in AEC teaching, cf. Fig.~\ref{fig:Concept}.

\section{Methods}  \label{sec:Methods}
This research employs a mixed-methods approach combining qualitative and quantitative evaluations methods similarly to previous design experiments.
%from \cite{johnson2014collaborative,fogarty2018improving}. 
In order to quantitatively assess the impact and usefulness of VR as a means of teaching construction details (Q1, Q2), a VR experience was developed and tested via examination in a pilot study. Acc. to Fig.~\ref{fig:FlowchartTesting}, participants first learn about a basement wall construction detail either with the VRE (test group) or with a 2D drawing (control group). Then they are tested on their newly acquired knowledge (exam). A final survey gathers information on the preexisting knowledge (before learning in the VRE) and infers socio-economic as well as professional background information together with participants perception of the VRE and potential future applications. As per today, no validated building design concept inventories exist, the example together with the test was developed from past teaching and exam problems to assess the understanding of the topic, cf. \ref{fig:Results_Tutorial_Model_Pics}. Further quantitative analysis upon the survey investigated students', instructors' but also AEC professionals' perceiving of the VR tool in current and future use in teaching and practice (Q3). Further details on the different methods used within this study are given in the remainder of this section.

\subsection{Elements of the Virtual Reality Experience}  \label{sec:VR_Experience}
\begin{figure} [h]
	\centering
	\includegraphics[width=0.55\linewidth]{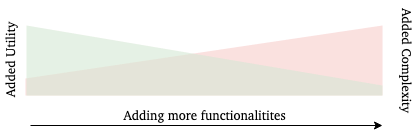} \\
	\caption{Utility and simplicity trade-off}
	\label{fig:TradeoffUtilitySimplicity}
\end{figure}

As this study pre-assumed a high degree of unfamiliarity with VR experiences amongst the panellists, the VRE itself was designed as simple and intuitive as possible. During the development process amongst the authors, it became clear that users could easily be overwhelmed with too many functionalities. The challenge was to find a balance between reducing complexity and maintaining completeness and usefulness. In order to make the VRE really intuitive, the number of functionalities in the app must be reduced as much as possible, while ensuring that the functionalities included play a significant role in providing the best learning outcomes for the users of the VRE. This trade-off is illustrated in Fig.~\ref{fig:TradeoffUtilitySimplicity}.

\subsubsection{Virtual Reality Environment}  \label{sec:VR_Experience_Env}
\begin{figure} [h]
	\centering
	\includegraphics[width=0.98\linewidth]{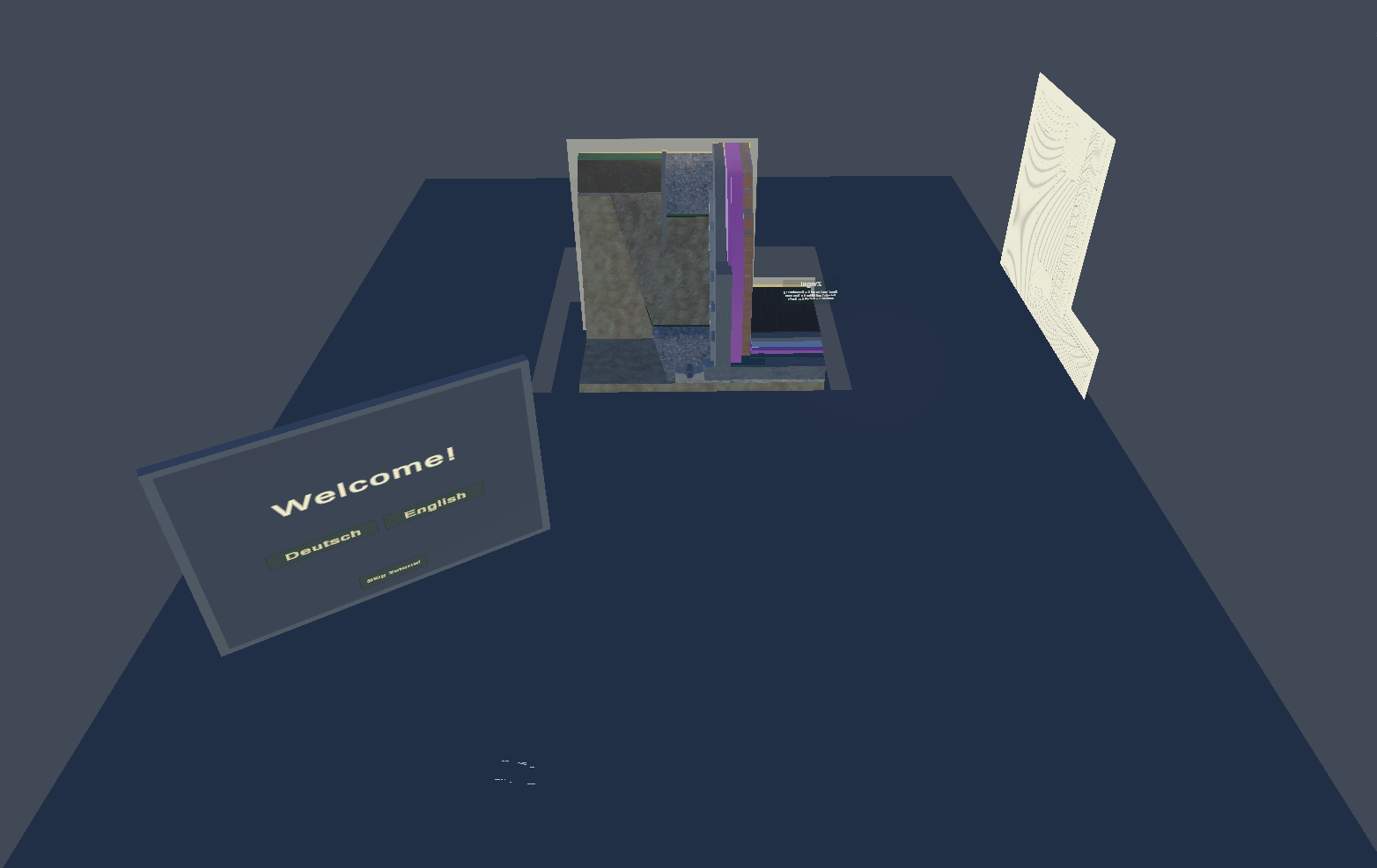} \\
	\caption{The VR-Environment with tutorial display, the 3D model and the 2D drawing (from left to right).}
	\label{fig:VREnvironement}
\end{figure}

The most fundamental part of the Virtual Reality Experience (VRE) is the environment the users find themselves in. Since humans are accustomed to being in rooms and are good at navigating them, the VRE is conceived as a medium sized room with the model of the construction detail in the centre. The user journey starts in one corner at the position of the screen for the tutorial. On the right hand side of the model as well as on one of the walls we provide a 2D drawing of the construction detail. The user is present in the scene yet without being represented by an avatar. Navigation is possible via a teleportation ray without the necessity to use portals or other distracting functionalities, cf. Fig.~\ref{fig:VREnvironement}. Users can see the controllers as well as a 'tablet' that displays information and hosts some of the functionalities. One of the most important functionalities is the so called quiz mode where users can playfully test their gained knowledge in the VRE.

\subsubsection{Controllers}  \label{sec:VRExp_Controllers}
While the user is not represented by an avatar in the VRE scene, the controllers are show in order to maintain support for orientation and immersion. However, the more important role of the controllers lies in enabling intuitive interactions with the model. The controllers in the VRE enable the user to see where the buttons on the controllers are without taking off the VR headset. We decided to keep the instructions hovering around the controller at all times in order to enable the users to quickly find the desired functionality without losing much time or attention.

\begin{figure} [h]
	\centering
	\includegraphics[width=0.4\linewidth]{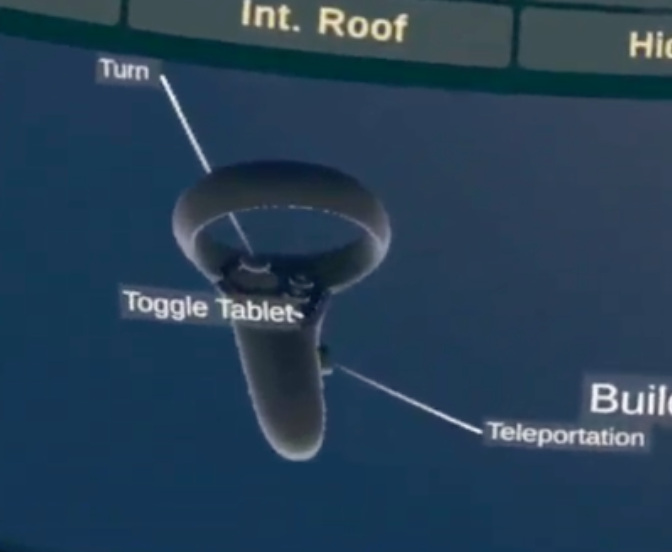} \\
	\caption{Excerpt from the tutorial: controller with labels indicating the functionality of some buttons.}
	\label{fig:VREnvironement_Controllers}
\end{figure}

\subsubsection{3D Model within the VRE}  \label{sec:VR_Experience_3DModel}
The model used as an example for this VRE is a section of a basement wall modelled after a 2D section from the book "Constructing Architecture" \cite{deplazes_constructing_2005}. It has been modelled in the CAD program Rhinoceros 3D. On one side, the section is straight, resembling a drawing. On the other side, parts of the model have been cut away, so that all elements remain visible, even if the whole model is shown.

\begin{figure} [h]
	\centering
	\includegraphics[width=0.98\linewidth]{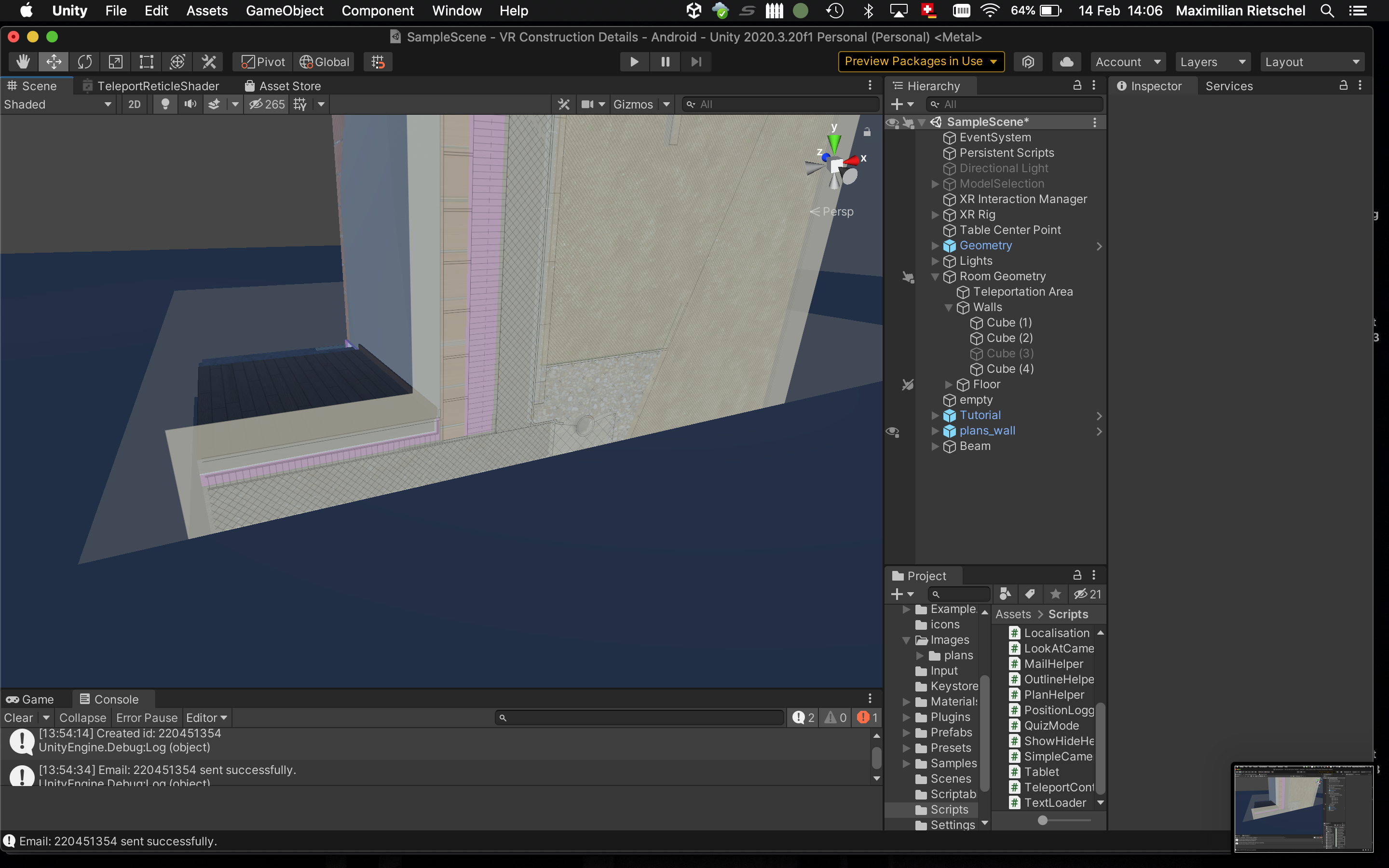} \\
	\caption{Excerpt from the 3D VR experience: model with drawing on the flat side.}
	\label{fig:VREnvironement_Drawings}
\end{figure}

Traditionally construction details are displayed as 2D sections. However, 3D models allow for much more information to be relayed in a much more intuitive way. We provide both, the 2D section together with the 3D model to foster the spacial and semantic understanding in one scene, cf. Fig.~\ref{fig:VREnvironement_Drawings}. In order to ensure inspection of the model from different angles easily, the controllers allow users to rotate and lift the model. With that, the model can be inspected closely at a fixed point within the VRE.

\begin{figure*}
	\centering
	\begin{tabular}{c c}
		\vspace{.5mm}
		\includegraphics[width=0.4\linewidth]{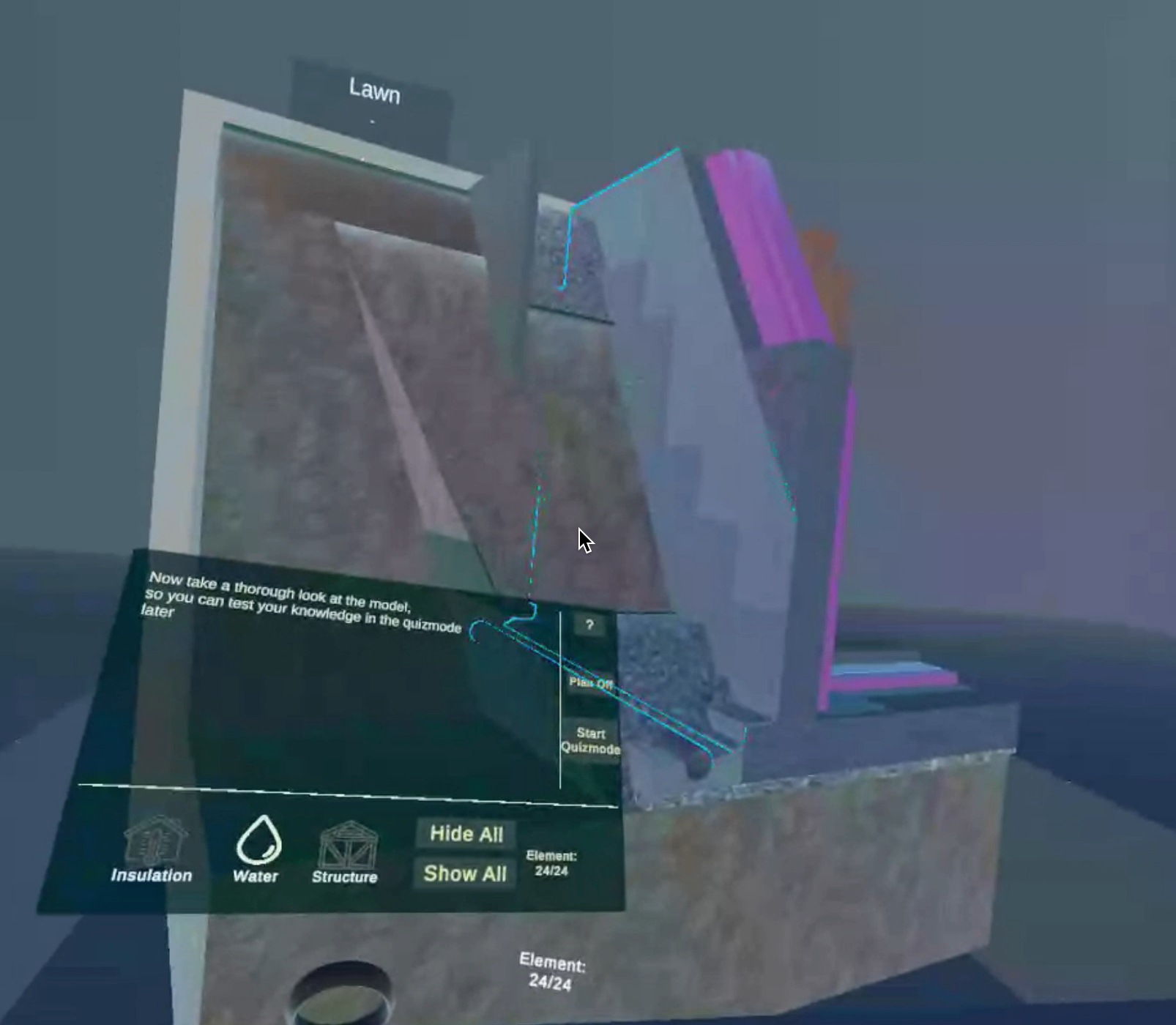} &
		\includegraphics[width=0.41\linewidth]{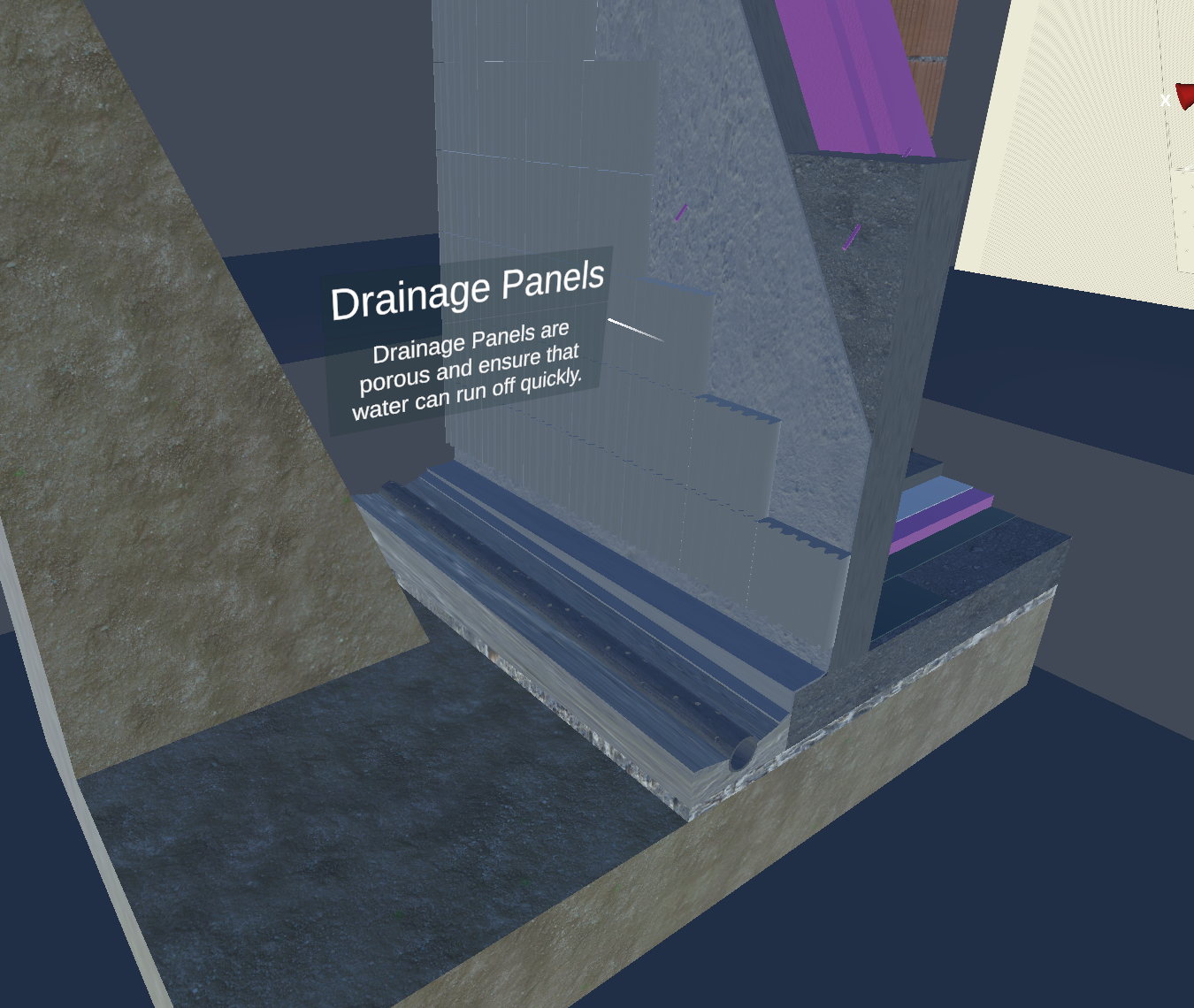} \\
	\end{tabular} 
	\caption{Examples from the 3D VR experience: (left) fully built up model, (right) example for an informative label.}
	\label{fig:Results_Tutorial_Model_Pics}
\end{figure*}

To convey the important information about the construction process of the detail, we implemented functionalities to assemble or dismantle the basement detail in an element-by-element fashion. For enabling a better understanding of the specific function of individual construction elements, cf. Fig.~\ref{fig:Results_Tutorial_Model_Pics}. Each element is equipped with a floating label with associated relevant information, including the name and a short explanation. When the user assembles the model step by step, the label for the newly enabled element is visible, resulting in a kind of guided walk-through of the construction process.

Finally, to support user's understanding of how some elements work together or which elements belong to certain construction groups, a functionality to highlight these groups of elements is provided. This function highlights a selected group (e.g. all elements of drainage or insulation) within the 3D model by colour.

\subsubsection{2D Drawings within the VRE}  \label{sec:VR_Experience_3DModel}
There are two 2D drawings in the VRE. One is placed on a wall in a certain distance to the 3D model. The intention here is to allow interested users an interaction with the tradition 2D sectional drawing of the detail. This drawing also includes labels, like the model itself, showing name and information about the elements.

Another 2D drawing is attached directly to the model. This drawing allows users to make a direct connection between the 2D sectional drawing and model. While it is not always clear from a 2D drawing what an element might represent, this presentation style is expected to clarify this easily by overlaying the 3D model. Users then immediately recognise elements from the 3D model on the 2D drawing.

\subsubsection{Tablet}  \label{sec:VR_Experience_Tablet}
\begin{figure} [h]
	\centering
	\includegraphics[width=0.7\linewidth]{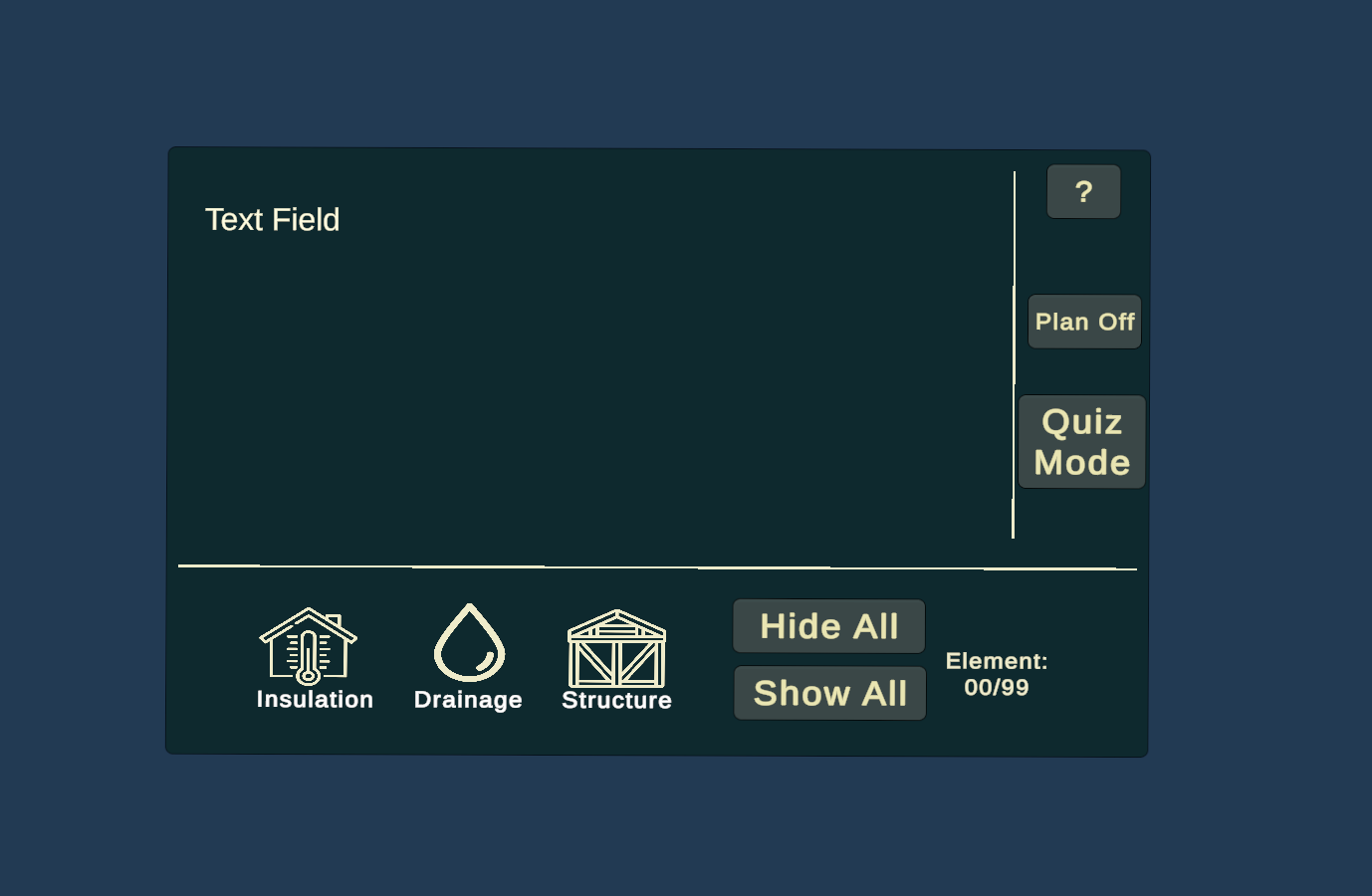} \\
	\caption{Excerpt from the 3D VR experience: overview of the tablet functionalities.}
	\label{fig:VREnvironement_Tablet}
\end{figure}

Most functionalities described so far can be accessed from the controllers, cf. Fig.~\ref{fig:VREnvironement_Controllers}. However, certain functionalities are are either not required often or might confuse the user by their complexity if located on the controller. Therefore, a screen - called the 'tablet' - hosts a number of buttons as well as a text field, cf. Fig.~\ref{fig:VREnvironement_Tablet}. The buttons enable interaction for the user with higher-level functionalities such as hiding or showing all elements at once or highlighting certain groups of elements or starting the quiz mode. As a means of communication between the app and the user, there is a text field on the tablet, where information, instructions or explanations can be relayed directly in a verbal way. Technically the tablet would allow for a more sophisticated development, including a catalogue of building elements with educational content such as photographs or even videos of the building elements in a real construction site. This is however not conducted within the scope of this study.

\subsubsection{Quizmode}  \label{sec:VR_Experience_Quizmode}
\begin{figure*}
	\centering
	\begin{tabular}{c c}
		\vspace{.5mm}
		\includegraphics[width=0.5\linewidth]{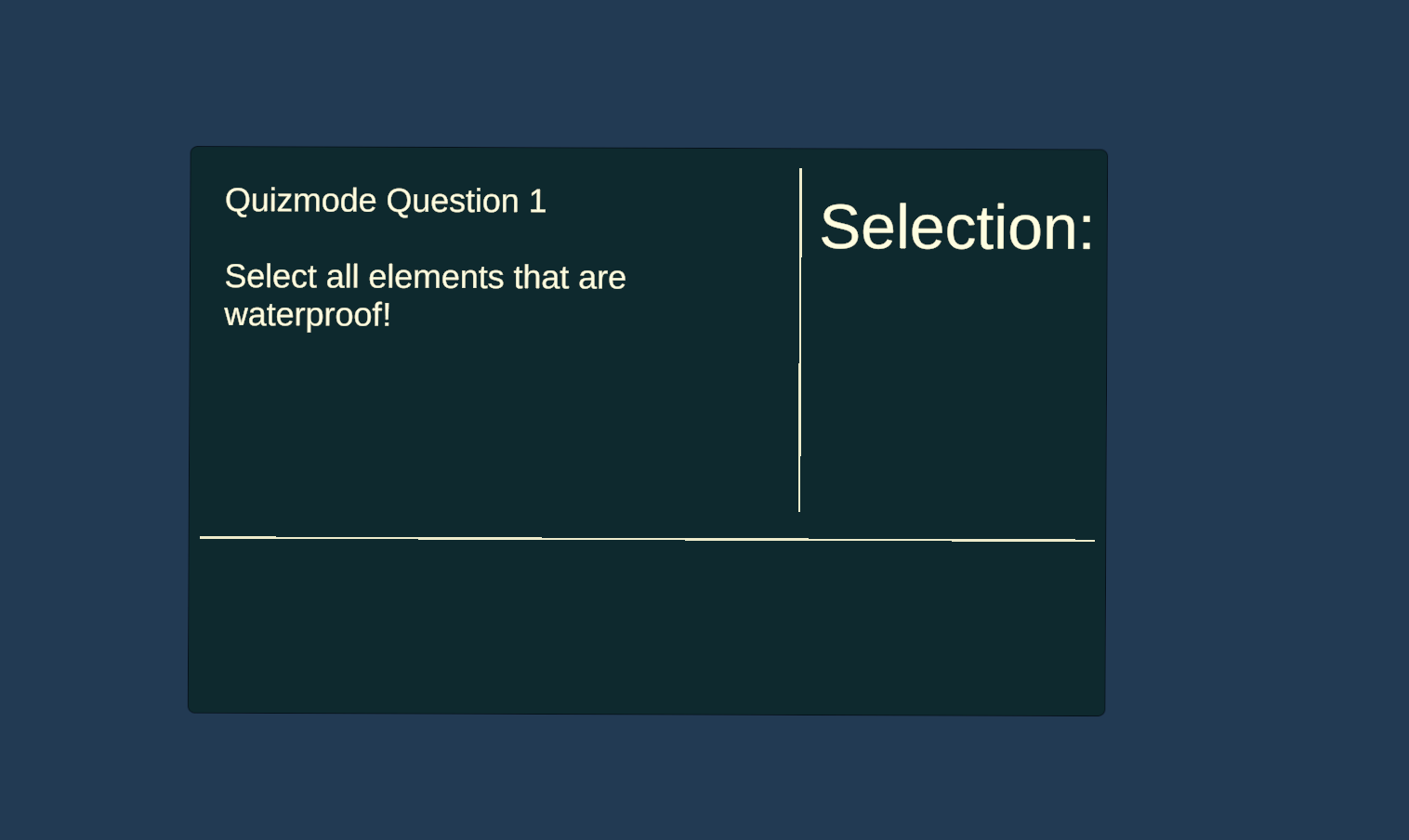} &
		\includegraphics[width=0.488\linewidth]{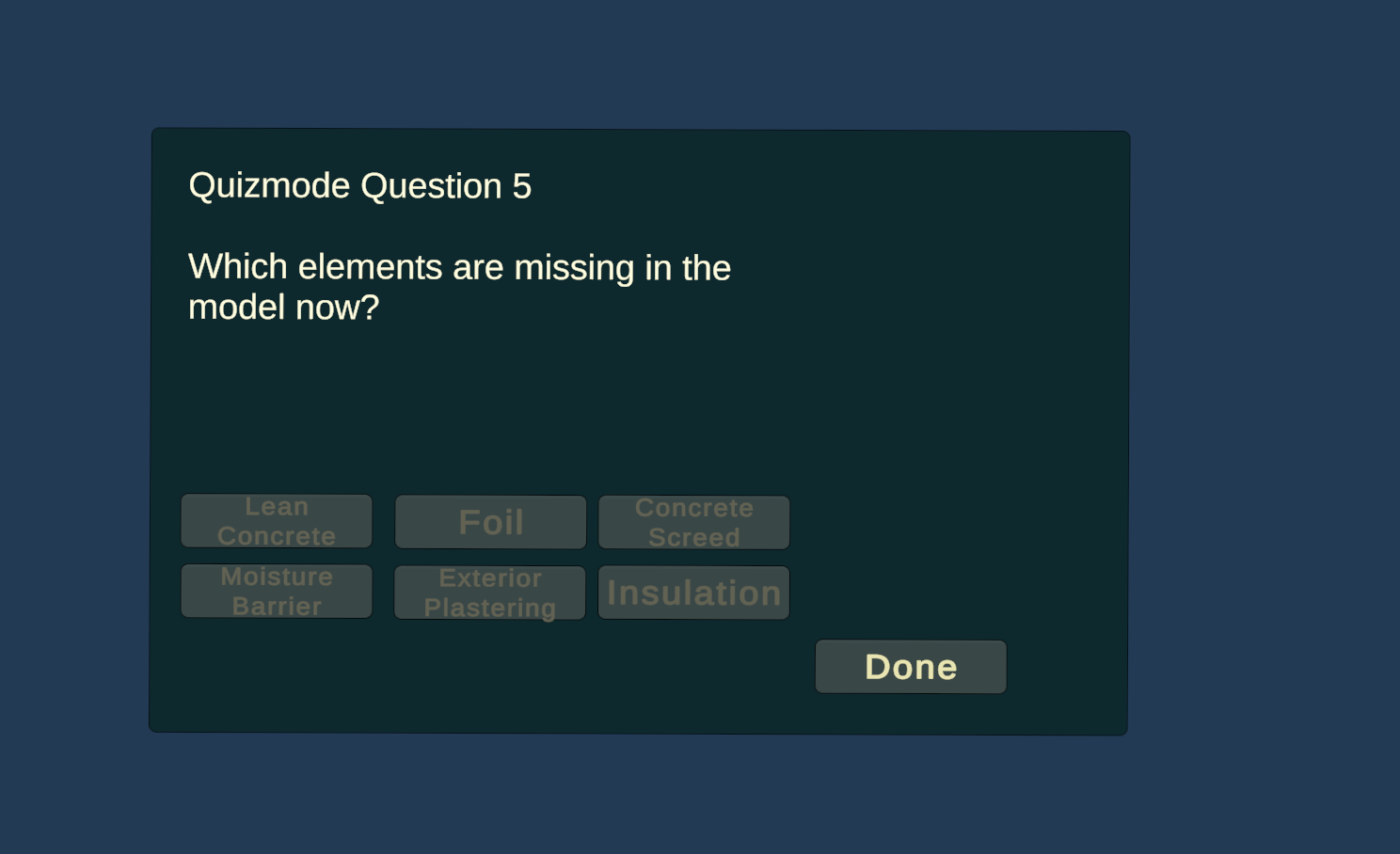} \\
	\end{tabular} 
	\caption{Excerpt from the 3D VR experience: example questions from the Quizmode for (left) selecting specific elements, and (right) answering in a multiple choice fashion.}
	\label{fig:Results_Tutorial_Model_QuizMode}
\end{figure*}

%As \cite{Maghool.2018} noted, important ways of learning for architectural education are interactive learning and problem-based learning. 
While the model is interactive, the interaction is one-sided only with the user changing what they see in the model. To make the learning experience more interactive, a so called Quizmode is introduced, cf. Fig.~\ref{fig:Results_Tutorial_Model_QuizMode}. The quiz mode brings a new dimension of learning into the VRE: the user is not merely looking at the model and adding or removing elements, but actually using their acquired knowledge to solve problems. The Quizmode presents the user with different types of problems to solve in form of questions. The first category are questions where the user has to select certain elements from the model, cf. Fig.~\ref{fig:Results_Tutorial_Model_QuizMode} (left). The second category of questions comprises different types of multiple choice questions, some related to how the elements work together, some testing knowledge about certain building elements and finally a question where a part of the model is missing and the user has to find out which elements are missing, cf. Fig.~\ref{fig:Results_Tutorial_Model_QuizMode} (right). By solving these problems, the user is tested towards gained understanding and correctness of their knowledge by receiving feedback. This study omits introduction of further interactive means within the VR for sake of simplicity while allowing problem solving. Further teaching methods such demanding users to select certain elements in the model or even change the model based on their knowledge could easily be implemented.

\subsubsection{Tutorial}  \label{sec:VR_Experience_Tutorial}
\begin{figure*} [h]
	\centering
	\begin{tabular}{c c c}
		\includegraphics[width=0.32\linewidth]{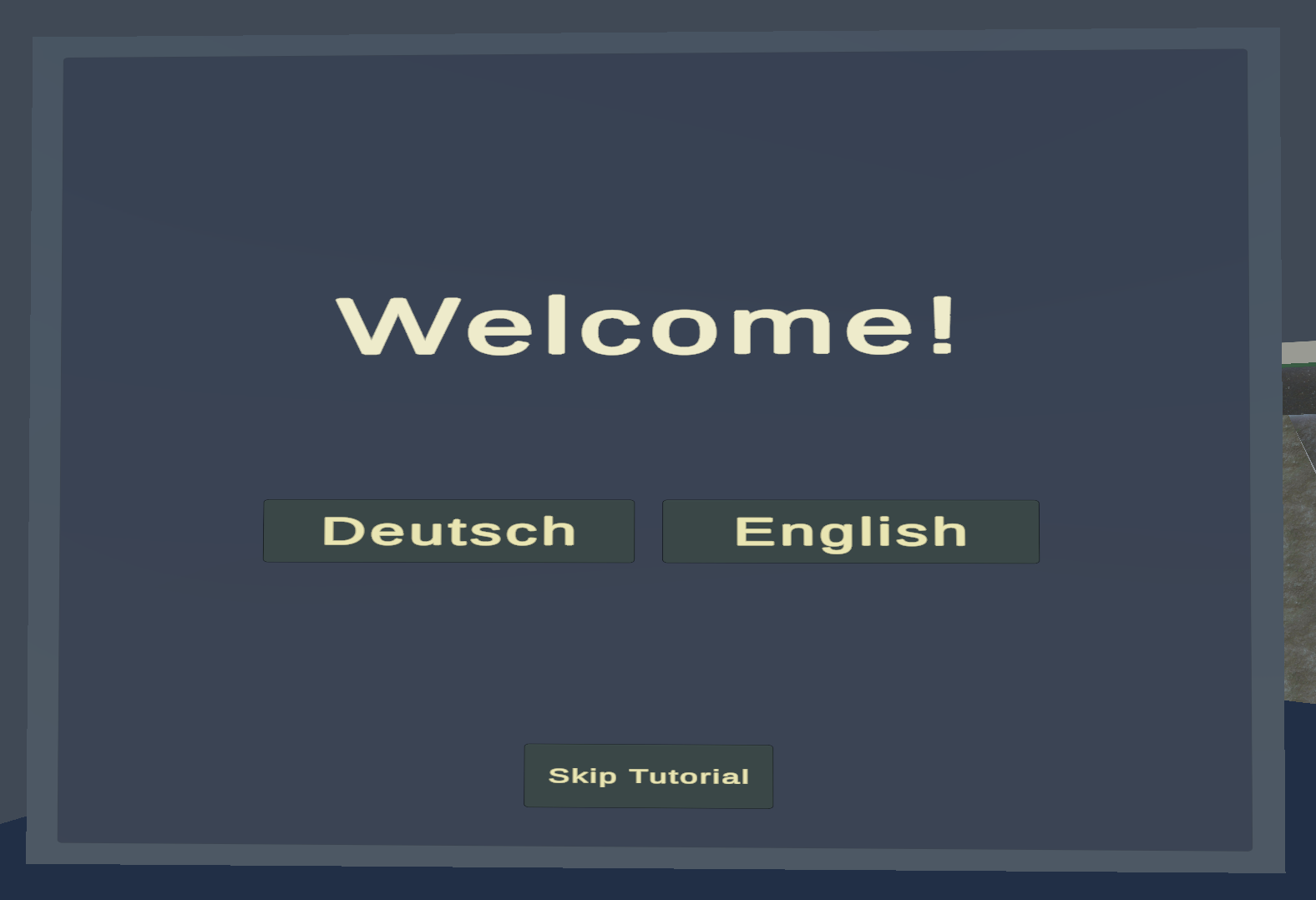} &
		\includegraphics[width=0.325\linewidth]{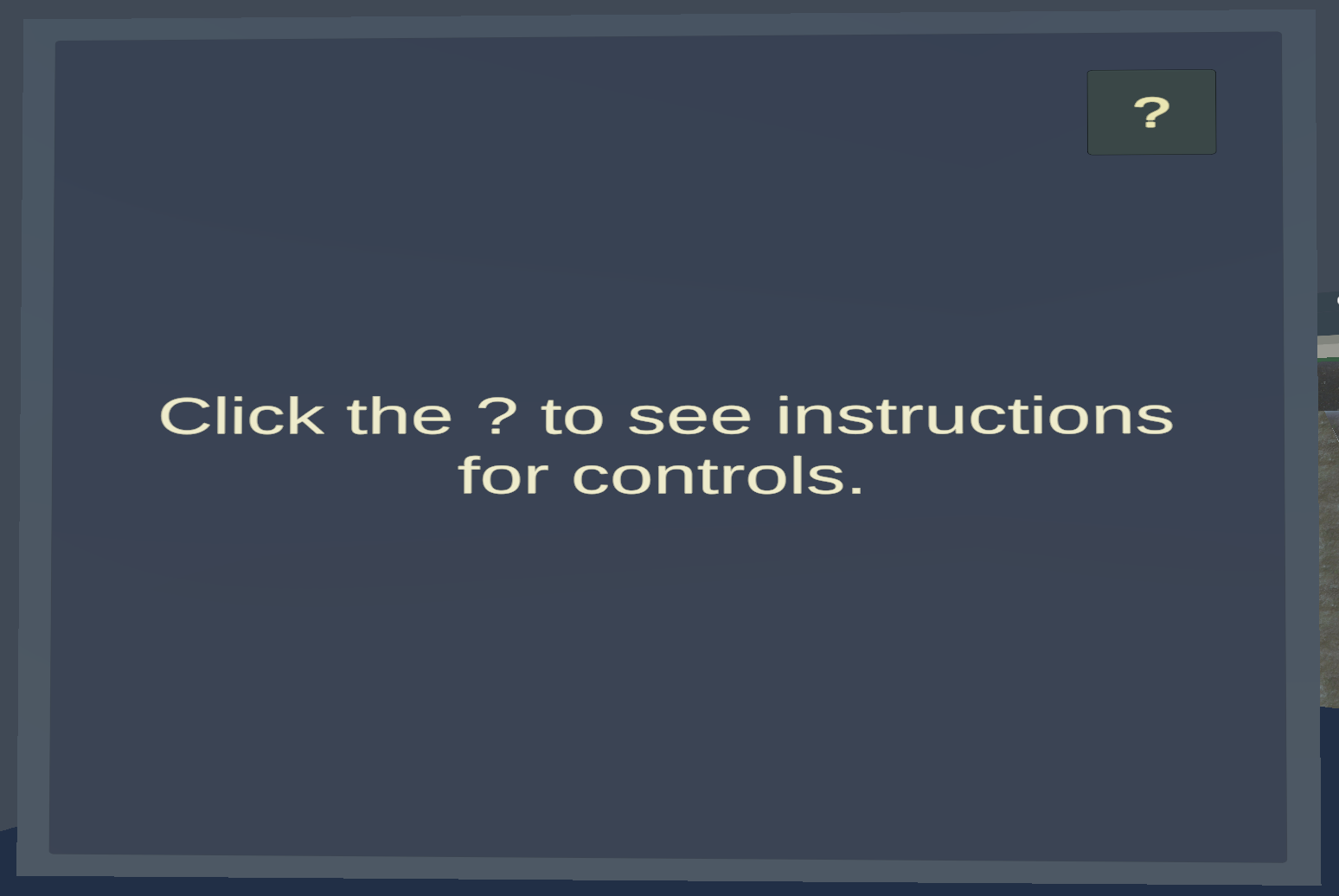} & 
		\includegraphics[width=0.32\linewidth]{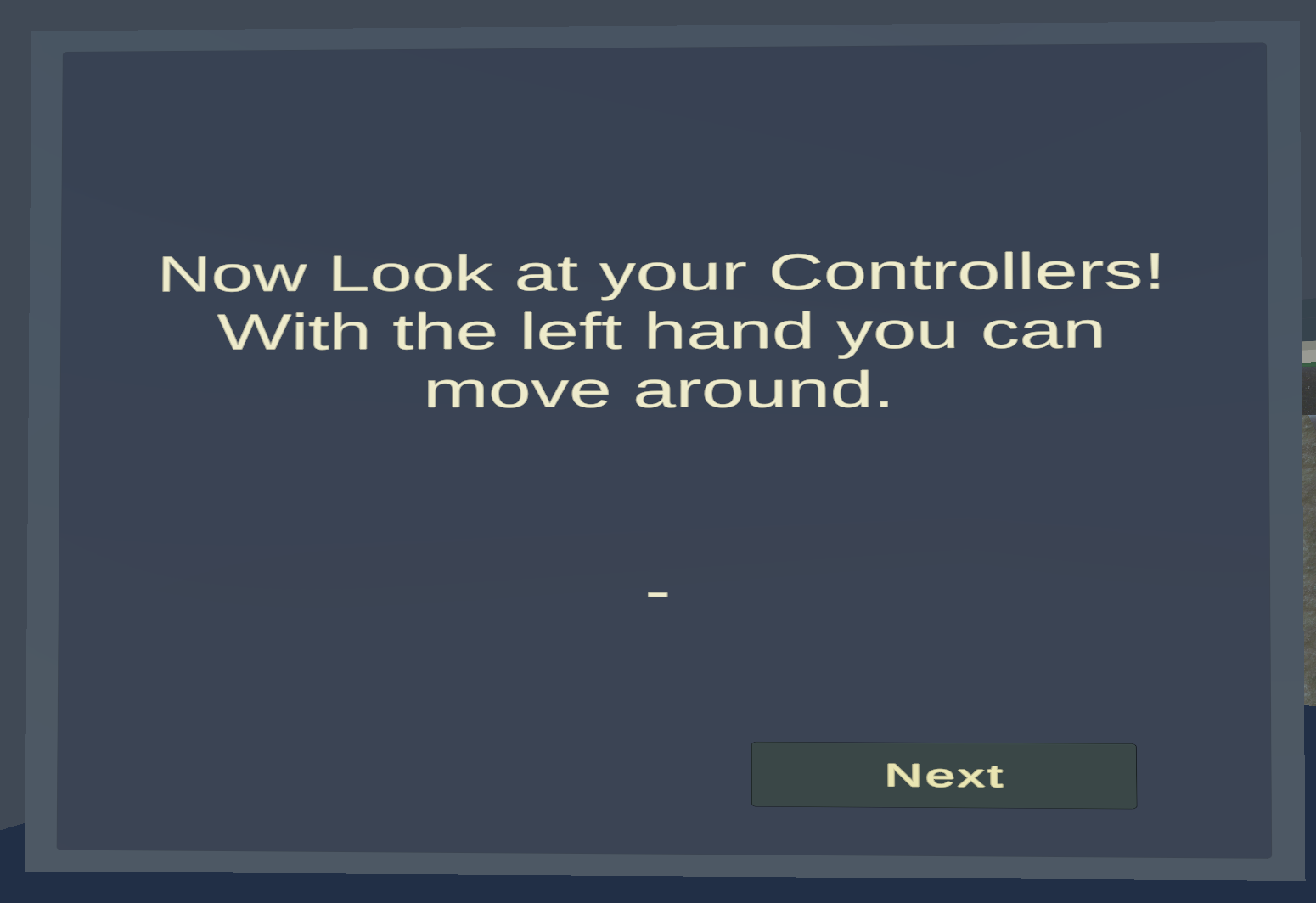} \\
	\end{tabular} 
	\caption{Excerpts form the Tutorial: (left) Welcome screen, (middle) starting instructions, (right) controller use instructions.}
	\label{fig:Results_Tutorial_Pics}
\end{figure*}

Since this study assumes many panellists being unfamiliar with VR experiences, a tutorial is provided. The implemented tutorial walks the users through the main functionalities in a concise manner without distracting users before finishing the tutorial, cf. Fig.~\ref{fig:Results_Tutorial_Pics}.

The tutorial starts with a welcoming screen, which allows the users to either chose the language (English or German) or skip the tutorial. The tutorial is of great importance, since it provides crucial information on how to navigate the VRE or interact with the content. Hence during the development we were guided by finding an optimal balance on providing relevant instructions in a concise manner while keeping this part short.

The tutorial is kept to a minimum number of lines of text. The user has the ability to review certain instructions with a help button found on the tablet, which is available throughout the whole experience. Using this button they can show and hide instruction labels that hover around the controllers and tablet and denominate certain functionalities on their corresponding buttons (Fig.~\ref{fig:Results_Tutorial_Pics}), enabling them to quickly find the appropriate buttons.

\subsubsection{Logs}  \label{sec:VR_Experience_Logs}
In order to allow future analyses of the users experiences and journeys within the VRE, a script in the background collects some user activities in a log file. Activities logged include the users orientation, buttons they press, changes in the model (hiding and showing elements), actions in the quiz mode and more. This information is collected in a text file along with the time and a unique ID for each session in the VRE. Despite the data is collected during this study, the evaluation and correlation to the exam scores is conducted in the future.

\subsection{Examining Effectiveness of the VRE}  \label{sec:VR_Examining}
Besides the formal assessment of the impact of the VRE onto the gained knowledge of the construction detail as described in the latter of this section, field notes were collected during observation of the panellists as they interacted with the 2D drawing as well as the different phases of the VRE experience. These observations were taken into consideration as qualitative information to interpret how the learning modes affect the panellists' understanding of the learning content and their interaction with it (Q1 and Q3).

\subsubsection{Participants and Data Collection}  \label{sec:VR_Examining_DataCollection}
The study panel consisted of 41 participants from within the Design++ network at ETH Zurich and is composed of:
\begin{itemize}
	\item 15 females and 26 males,
	\item 12 resp. 29 persons speaking English resp. German
	\item \{17, 22, 2\} persons in the age range of \{(18-24), (25-30), (31-40)\},
	\item \{11, 17, 13\} persons holding a \{High School, Bachelor, Master\} degree,
	\item \{20, 3, 18\} persons with a background in \{architecture, civil engineering, others\},
	\item \{3, 18, 6, 3, 11\} persons with an amount of enrolled semesters \{(1-2), (3-6), (7-10), (>11), finished\},
\end{itemize}

\begin{figure} [h]
	\centering
	\includegraphics[width=0.4\linewidth]{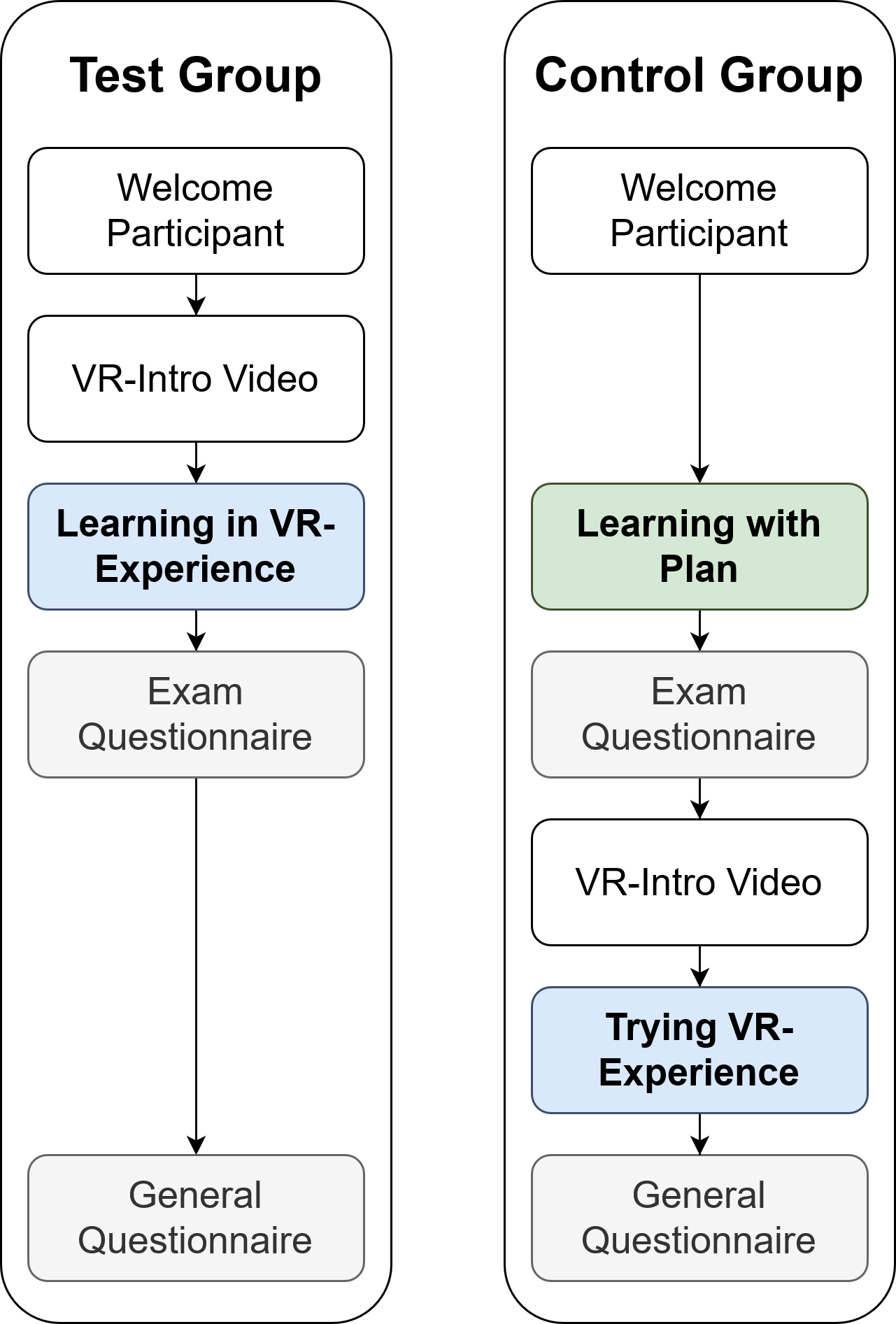} \\
	\caption{Flow chart of the process participants go through in this study.}
	\label{fig:FlowchartTesting}
\end{figure}

Given the consistence of the panel, participants with a wide range and diversity of understanding chose to participate and experience the VR models. The panel was randomly divided into a test group and a control group. They then learned with the help of either the VRE (test group) or with a 2D drawing (control group). After this learning period, both groups filled out an ‘exam’ where their knowledge was tested. After the exam, participants in the control group had the opportunity to test the VRE as well. The last part for both groups was to fill out a survey where they supply information about their background and their opinions on VR for learning. The whole process is illustrated in Fig.~\ref{fig:FlowchartTesting}.

\subsubsection{Learning in the VRE}  \label{sec:VR_Examining_VRELearning}
For learning with the VRE, the participants in the test group are shown an introductory video briefly introducing the VRE. After watching the video, the participants are instructed on how to use the VR headset and the controllers. Then they enter the VRE and conduct the tutorial. Subsequently, the users are granted time to inspect the model and learn on their own pace. Whenever the panellist feels ready, but no earlier than 3 minutes into the VRE, they can start the Quizmode to test their knowledge. After finishing the Quizmode, the participants are free to further browse and explore the model until satisfied.

\subsubsection{Learning with 2D drawings}  \label{sec:VR_Examining_2DPlans}
Participants from the control group are given two sheets of paper (provided as supplemental material, cf. Sec.~\ref{sec:Supplemental}). The first sheet contains the a 2D section drawing of the basement construction detail, which is exactly the one to be found in the VRE, cf. Fig.~\ref{fig:VREnvironement_Drawings}. The building elements in the drwaing are labelled and the wall- and floor- construction are listed separately on the top of the page. The second sheet contains a table with short explanations for each building element, which again are identically the same explanations as on the labels within the VRE. The participants receive verbal instructions on how to memorise the information on the 2D drawing. Subsequently, the test group participants are granted time learn on their own pace until they feel ready to take the exam.

\subsubsection{Exam}  \label{sec:VR_Examining_Exam}
After the learning phase, an exam is testing the participants’ knowledge. The exam consists of thirty questions of two types. The first type is a simple true/false question, where participants have to assess whether a given statement towards the construction detail is true or false. The questions can be about the location of certain elements (e.g. "Between the concrete screed and the insulation there must be a separating foil." [true]), about the sequence of the building process (e.g. "Before applying the exterior plastering, the drainage panels need to be mounted." [false]) or about characteristics of certain elements (e.g. "The exterior plastering does not need to be watertight, because the concrete wall already is." [false]).

The second type is a multiple-choice question, where the participants have to select all correct answers from a list of possible answers. The questions are again covering the aspects of location in the construction (e.g. "Which of these elements are within the insulation layer (warm)?"), about the sequence of construction (e.g. "Which of these elements are added before the floor insulation is built in?"), or about the characteristics of certain elements (e.g. "Which of these elements are waterproof?"). 

Each question is rewarded with one point if answered correctly and 0 points otherwise. There is no time limit imposed for the conduction of filling the exam.

\subsubsection{Survey}  \label{sec:VR_Examining_Survey}
Surveys and interviews as described in this section were used to assess student and instructor perception of the VR tools (Q3). The survey questionnaire is provided as supplemental material, cf. Sec.~\ref{sec:Supplemental}, and starts by inferring about the participants background, including questions towards age, profession and education. Then, a set of questions aimed at assessing the participants preexisting knowledge about construction details, their familiarity with VR and their experience in the VRE as well as their general opinion on VR for learning is following. For these questions, the panellists are presented statements to either agree or disagree on a five-point Likert scale (ranging in "strongly disagree", "disagree", "neutral", "agree" and "strongly agree"). These answers are then transformed into a score ranging form 0 to 1, where "strongly disagree" maps to 0 and "strongly agree" maps to 1. 

The pre-test for pre-existing knowledge is highly important, since preexisting knowledge could significantly influence the participants' performance in the study. Here the participants are asked to rate their familiarity with construction details (such as "In general, I'm familiar with construction details"). It is also assessed, whether the information in the study would be new to the participant (e.g. "I would have known most of the information about the construction detail without the learning part of this study anyways."). Further information is provided in Sec.~\ref{sec:VR_Examining_sPEK}.

Next, the familiarity of the panellist with VR is assessed in a similar fashion (e.g. "I have used VR apps (including games) often."). Subsequent questions aim at gathering participants' feedback about the VR application, covering user-friendliness of the VRE (e.g. "I quickly learned how to navigate the app.") or the usefulness of functionalities (e.g. "The labels with short explanations of each element are useful.").

The last section asks participants about their opinion on VR for learning and teaching in general via rather general statements (e.g. "Using VR improves learning."), statements aiming at the enjoyment of the experience (e.g. "Learning with a VR App is fun."), or statements to assess whether the participant would learn with VR again (e.g. "I would like to learn with VR in other areas as well.").

\subsection{Measuring the Learning Outcome}  \label{sec:VR_Examining_LearningOutcome}
In order to assess whether the VRE provides a better learning experience and outcome than the traditional way of learning about construction details via 2D sectional drawings, three measures need to be established: i) the pre-existing knowledge score $s_{PEK}$, ii) the exam score $s_{exam}$, and iii) the knowledge gain $\partial_{knowledge}$ via the method of study.

\subsubsection{Exam Score $s_{exam}$}  \label{sec:VR_Examining_Sexam}
The exam-score represents how well the participant performed on the exam. Since almost all of the exam questions are simple true/false questions, someone filling it out randomly would still receive 50\% of the possible points on average. Therefore, we suggest to compute $s_{exam}$ via:
\begin{equation}    \label{eq:Sexam}
	s_{exam} = \frac{n_{ex,correct} - 0.5 \cdot n_{ex,total}}{0.5 \cdot n_{ex,total}} \;.
\end{equation}
where $n_{ex,correct}$ is the number of exam questions answered correctly and $n_{ex,total}$ is the total number of question in the exam. Half of the maximum score is deducted from the actual score to account for the mentioned bias. The remaining number of points is then mapped to the interval between 0 and 1 by division by half the maximum score.

For some investigations, we evaluate the average (i.e. the expected or mean) exam score $<s_{exam}>$ as:
\begin{equation}    \label{eq:Sexam}
	<s_{exam}> = \frac{n_{ex,correct}}{n_{ex,total}} \;.
\end{equation}

\subsubsection{Pre-existing Knowledge Score $s_{PEK}$}  \label{sec:VR_Examining_sPEK}
The pre-existing knowledge is assessed by means of 8 questions during the survey as already outlined in Sec.~\ref{sec:VR_Examining_Survey} using a five-point Likert scale, cf. Sec.~\ref{sec:VR_Examining_Survey}. 
The pre-existing knowledge (PEK) score $s_{PEK}$ is computed as the mean of all answers and translates the five-point Likert scale into the interval from 0 (no preexisting knowledge) to 1 (good knowledge).

\subsubsection{Knowledge Gain $\partial_{knowledge}$}  \label{sec:VR_Examining_deltaKnowledge}
The comparison of the score for pre-existing knowledge ($s_{PEK}$) with the score achieved in the exam ($s_{exam}$) is conducted in order to establish a measure for the learning outcome and knowledge gain $\partial_{knowledge}$. However, establishing a mathematical formula is ambiguous and discussed for two approaches in the latter of this section. 

For both measures, Welch's t-tests were used to assess the statistical significance of the difference in mean values of the test and control group results for  determining the influence of the VR experience. Welch's t-tests assume data to be obtained from an independent, random sample of panellists learning about the basement design detail and that the distribution of scores within each of the two compared groups is normal but the variances of the two groups may be unequal.

\paragraph{Method 1 - non linear}
Measure method 1 creates a relationship between $s_{exam}$ and $s_{PEK}$ by dividing the exam-score by the PEK-score. As the PEK-score in the denominator can be zero, we add 1 to denominator to avoid division by zero and as well the nominator for consistency. A factor of 0.5 ensure a maximum score of 1, resulting in the following function for $\partial_{knowledge}$ acc. to method 1:
\begin{equation}    \label{eq:partialKnowl_m1}
	\partial_{knowledge} = 0.5 \cdot \frac{1 + s_{exam}}{1 + s_{PEK}} \;.
\end{equation}

The possible outcomes for the extreme values of $s_{exam}$ and $s_{PEK}$ are provided in Tab.~\ref{tab:dknow_Method_1}).

\begin{table}[]
	\centering
	\caption{$\partial_{knowledge}$ for Method 1}
	\label{tab:dknow_Method_1}
	\begin{tabular}{ccc}
		\hline\hline
		\multirow{2}{*}{$s_{PEK}$} & \multicolumn{2}{c}{$s_{exam}$} \\  
		& \multicolumn{1}{c}{0}    & 1    \\ 
		0                     & \multicolumn{1}{c}{0.50} & 1.00 \\ 
		1                     & \multicolumn{1}{c}{0.25} & 0.50 \\ 
		\hline\hline
	\end{tabular}
\end{table}

With this method, equal scores $s_{PEK}$ and $s_{exam}$ always yield a $\partial_{knowledge}$ of 0.5, meaning the person did not learn anything. With $s_{exam}$ being higher than $s_{PEK}$ the result will be between 0.5 and 1 to indicate that a person learned something. Otherwise, $\partial_{knowledge}$ will be between 0.25 and 0.5. In that case the participants rated their pre-existing knowledge better than how they performed in the exam.

\paragraph{Method 2 - linear}
The relationship between $s_{PEK}$, $s_{exam}$ and $\partial_{knowledge}$ in Method 1 is not linear. However, we can design a formula for $\partial_{knowledge}$ that is linear and maps the learning outcome to an interval spanning between -1 to 1, cf. Tab.~\ref{tab:dknow_Method_2}.
\begin{equation}    \label{eq:partialKnowl_m2}
	\partial_{knowledge} = s_{exam} - s_{PEK}
\end{equation} 

\begin{table}[]
	\centering
	\caption{$\partial_{knowledge}$ for Method 2}
	\label{tab:dknow_Method_2}
	\begin{tabular}{ccc}
		\hline\hline
		\multirow{2}{*}{$s_{PEK}$} & \multicolumn{2}{c}{$s_{exam}$}       \\   
		& \multicolumn{1}{c}{0}    & 1    \\  
		0                     & \multicolumn{1}{c}{0.00} & 1.00 \\  
		1                     & \multicolumn{1}{c}{-1.00} & 0.00 \\ 
		\hline\hline
	\end{tabular}
\end{table}

This relationship is linear and the value for equal scores $s_{PEK}$ and $s_{exam}$ (no learning outcome) lies at $\partial_{knowledge} = 0$. Values between 0 and 1 indicate a positive learning outcome with knowledge gain. 

\section{Results} \label{sec:Results}
The development of the VRE was conducted in three cycles, where a first phase conducted interviews amongst the authors towards the selection of construction details and rough code implementations. Based on the judgement of finding a good balance between complexity and appropriateness of the construction details for an undergraduate teaching scenario as well as the efforts of the software development, an iterative coding process was implemented for the basement construction detail as shown in Fig.~\ref{fig:VREnvironement} and described in Sec.~\ref{sec:VR_Experience}. The final implementation of the VRE was then used to conduct the experiment as described in Sec.~\ref{sec:VR_Examining}. The obtained results of the experiment are provided in the latter of this section.

%\subsubsection{Learning outcomes}  \label{sec:Results_Study_LearningOutcomes}
As outlined in Sec.~\ref{sec:VR_Examining_DataCollection}, the panel of this study consisted of 41 participants, where 23 had a background in architecture or civil engineering (considered "Professionals") while the remaining 18 came from various other backgrounds (considered "Other"), cf. Sec.~\ref{sec:VR_Examining_DataCollection}. 

Tab.~\ref{tab:StudyResMeanVals} summarises the main results of this study for the mean values of the scores $s_{PEK}$ and $s_{exam}$ together with the survey ratings for prior VR familiarity ("VR PEK Score"), user experience within our VRE ("VR UX Score"), enjoyment of our VRE ("VR Liking Score") and their perception of VR in education ("VR in EDU Score"). The quantity "VR Quiz Score" reports the average score of points obtained in the quiz mode for those panellists, who have studied the construction detail with the 2D drawing on paper first (control group) and were then exposed to VR. The outcomes are reported for all participants, conditioned on being either in the test or control group, and conditioned on the different backgrounds. This allows for subsequent hypothesis testing with different conditions.

\begin{table}[]
	\centering
	\caption{Mean values for the most relevant variables}
	\label{tab:StudyResMeanVals}
	\begin{tabular}{lccccccc}
		\hline\hline
		&
		\textbf{All} &
		\textbf{Test} &
		\textbf{Control} &
		\textbf{\begin{tabular}[c]{@{}c@{}}Test\\ Professional\end{tabular}} &
		\textbf{\begin{tabular}[c]{@{}c@{}}Control\\ Professional\end{tabular}} &
		\textbf{\begin{tabular}[c]{@{}c@{}}Test\\ Other\end{tabular}} &
		\textbf{\begin{tabular}[c]{@{}c@{}}Control\\ Other\end{tabular}} \\ 
		$N_{participants}$   & 41     & 22    & 19     & 11     & 12     & 11    & 7      \\ 
		$s_{exam}$  [-]           & 0.726  & 0.734 & 0.715  & 0.769  & 0.820  & 0.699 & 0.551  \\ 
		$s_{PEK}$  [-]          & 0.471  & 0.384 & 0.572  & 0.622  & 0.727  & 0.145 & 0.308  \\ 
		$\partial_{knowledge,M1}$    & 0.485  & 0.519 & 0.445  & 0.454  & 0.450  & 0.584 & 0.436  \\ 
		$\partial_{knowledge,M2}$    & -0.072 & 0.018 & -0.182 & -0.149 & -0.188 & 0.185 & -0.174 \\ 
		VR Quiz Score [-]   & 0.666  & 0.619 & 0.740  & 0.659  & 0.750  & 0.591 & 0.722  \\ 
		VR PEK Score [-]         & 0.372  & 0.466 & 0.263  & 0.420  & 0.271  & 0.511 & 0.250  \\ 
		VR UX Score [-]          & 0.683  & 0.662 & 0.707  & 0.580  & 0.755  & 0.744 & 0.625  \\ 
		VR Liking Score [-]      & 0.693  & 0.684 & 0.703  & 0.652  & 0.708  & 0.717 & 0.694  \\ 
		VR in EDU Score [-] & 0.726  & 0.731 & 0.719  & 0.674  & 0.722  & 0.788 & 0.714  \\ 
		Time in VR [s]      & 1001   & 1028  & 959    & 952    & 945    & 1081  & 983    \\
		\hline\hline
	\end{tabular}
\end{table}

The Welch's version of the student t-test has been performed on selected study variables to investigate whether there is a significant difference between the test group and the control group. The investigated variables are: $\partial_{knowledge,M1}$ and $\partial_{knowledge,M2}$ for methods 1 and 2, $s_{exam}$, the score for the familiarity with VR technology ("VR PEK") and the score for how easy it was for the participants to navigate the app ("VR UX").

\begin{table}[]
	\centering
	\caption{p values for the most relevant variables}
	\label{tab:StudyResttestVals}
	\begin{tabular}{lcccccc}
		\hline\hline
		& ${p_{All}}$ & ${p_{Others}}$ & ${p_{Prof}}$ & ${tstat_{All}}$ & ${tstat_{Others}}$ & ${tstat_{Prof}}$ \\
		$\partial_{knowledge,M1}$ & 0.0209 & 0.0010 & 0.9161 & 2.4105  & 3.9989  & 0.1067  \\ 
		$\partial_{knowledge,M2}$ & 0.0328 & 0.0006 & 0.7750 & 2.2154  & 4.2913  & 0.2897  \\ 
		$s_{PEK}$         & 0.0383 & 0.0758 & 0.1378 & -2.1441 & -1.8988 & -1.5428 \\ 
		$s_{exam}$        & 0.9791 & 0.0760 & 0.4809 & 0.0263  & 1.8973  & -0.7182 \\ 
		VR PEK        & 0.0936 & 0.2044 & 0.3388 & 1.7186  & 1.3232  & 0.9789  \\ 
		VR UX        & 0.4613 & 0.1989 & 0.0323 & -0.7440 & 1.3401  & -2.2928 \\ 
		\hline\hline
	\end{tabular}
\end{table}

\subsubsection{Results for Research Question Q1: Does VR improve student's understanding of complex spatial construction arrangements in contrast to 2D drawings?}  \label{sec:Results_Study_LearningOutcomes_Q1}

\begin{figure*}
	\centering
	\begin{tabular}{c c}
		\vspace{.5mm}
		\includegraphics[width=0.45\linewidth]{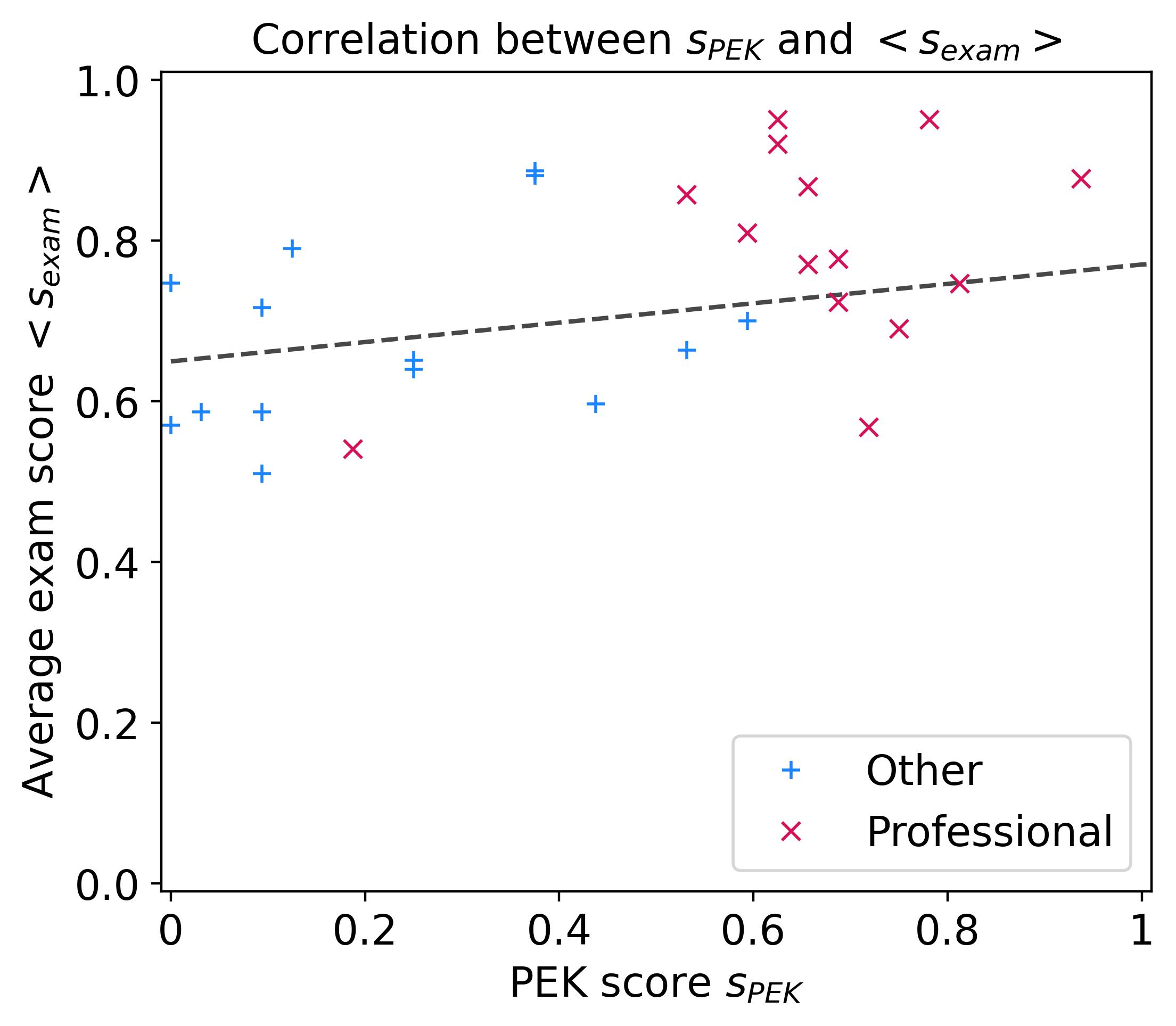} &
		\includegraphics[width=0.45\linewidth]{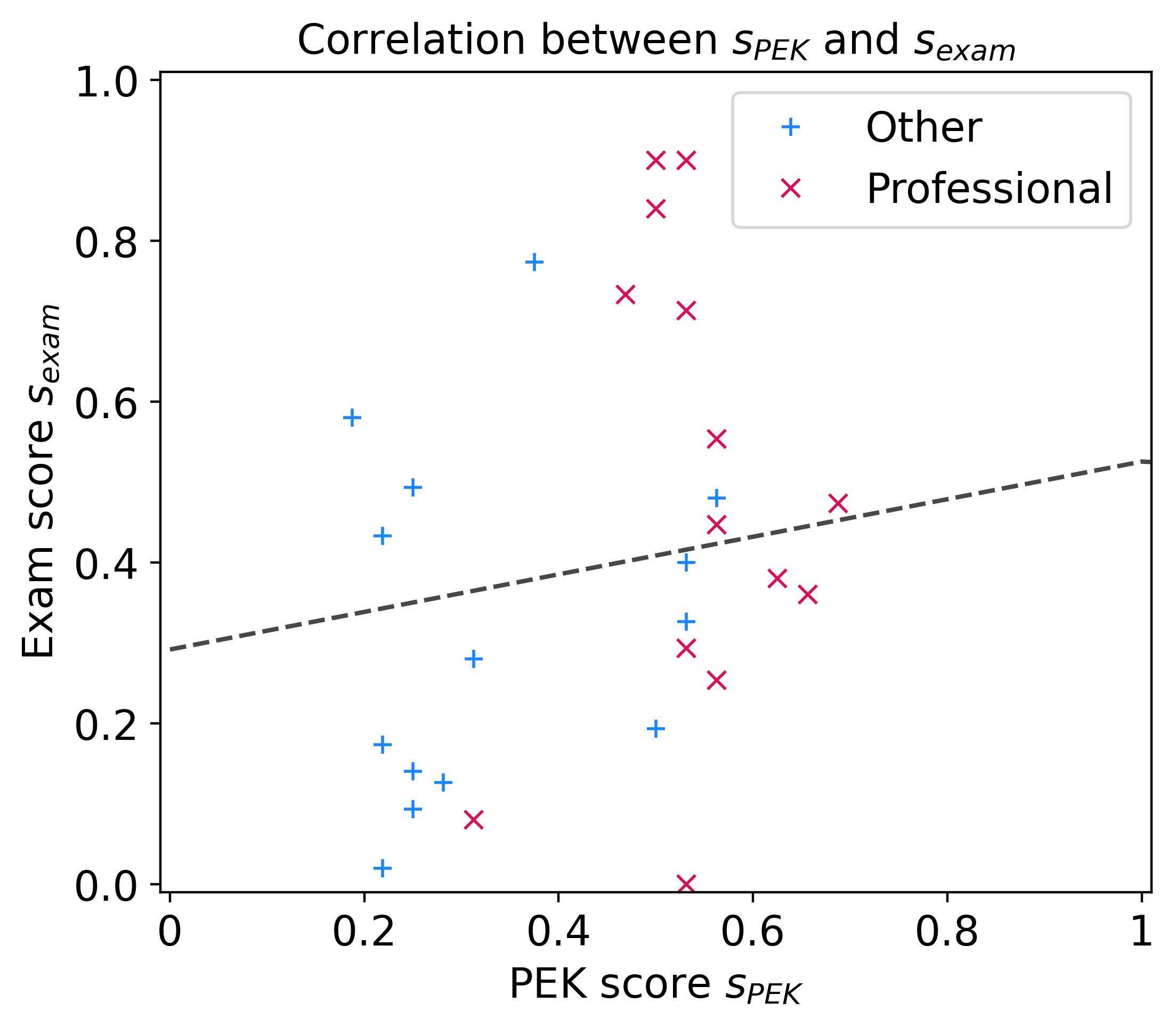} \\
	\end{tabular}
	\caption{Correlation graphs between PEK score $s_{PEK}$ and: (left) average score; (right) exam score $s_{exam}$}
	\label{fig:Results_AvgScore_Corr}
\end{figure*}

Inspection of Tab.~\ref{tab:StudyResMeanVals} as well as Fig.~\ref{fig:Results_AvgScore_Corr} shows, that while test group and control group performed very similarly on the exam, the professionals performed better on the exam than the others. Surprisingly, among professionals, $s_{exam}$ is higher in the control group than in the test group. However, $s_{PEK}$ is also higher in the control group amongst professionals. While $\partial_{knowledge}$ is similar for the groups amongst professionals, for participants from other backgrounds the values differ significantly. The control group scored higher on the VR Quiz, which is not surprising because they have learned the detail on paper and filled out an exam before they took the VR quiz. The VRE user experience and the liking of the VRE and whether VR was deemed useful for education were all rated similarly amongst the different groups. The time spent in the VRE was, slightly higher for the test group.

\begin{figure} [h]
	\centering
	\includegraphics[width=0.57\linewidth]{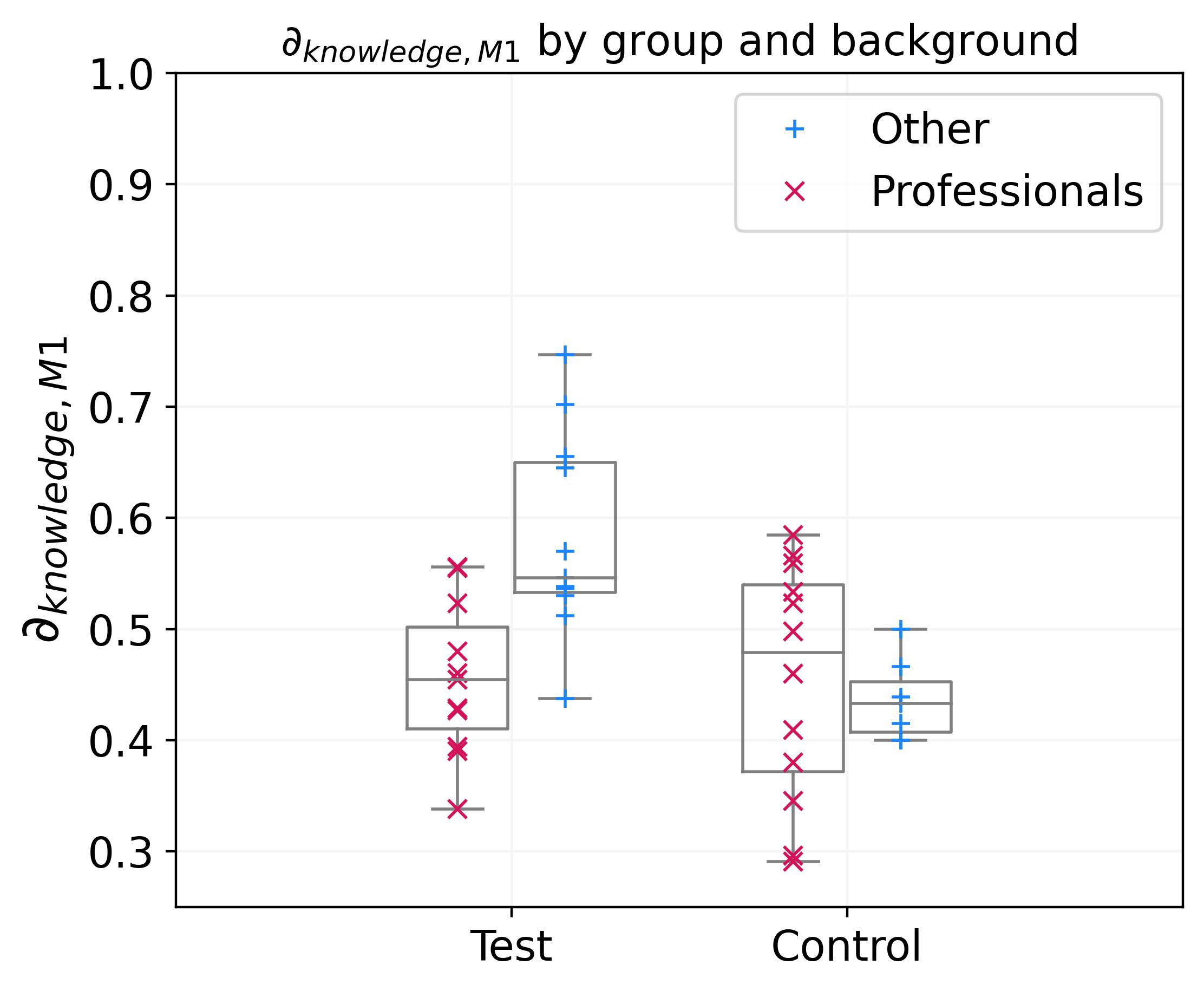} \\
	\caption{Gained knowledge $\partial_{knowledge}$ by group and background (Method 1).}
	\label{fig:Results_Knowledgegain_M1}
\end{figure}

\begin{figure} [h]
	\centering
	\includegraphics[width=0.75\linewidth]{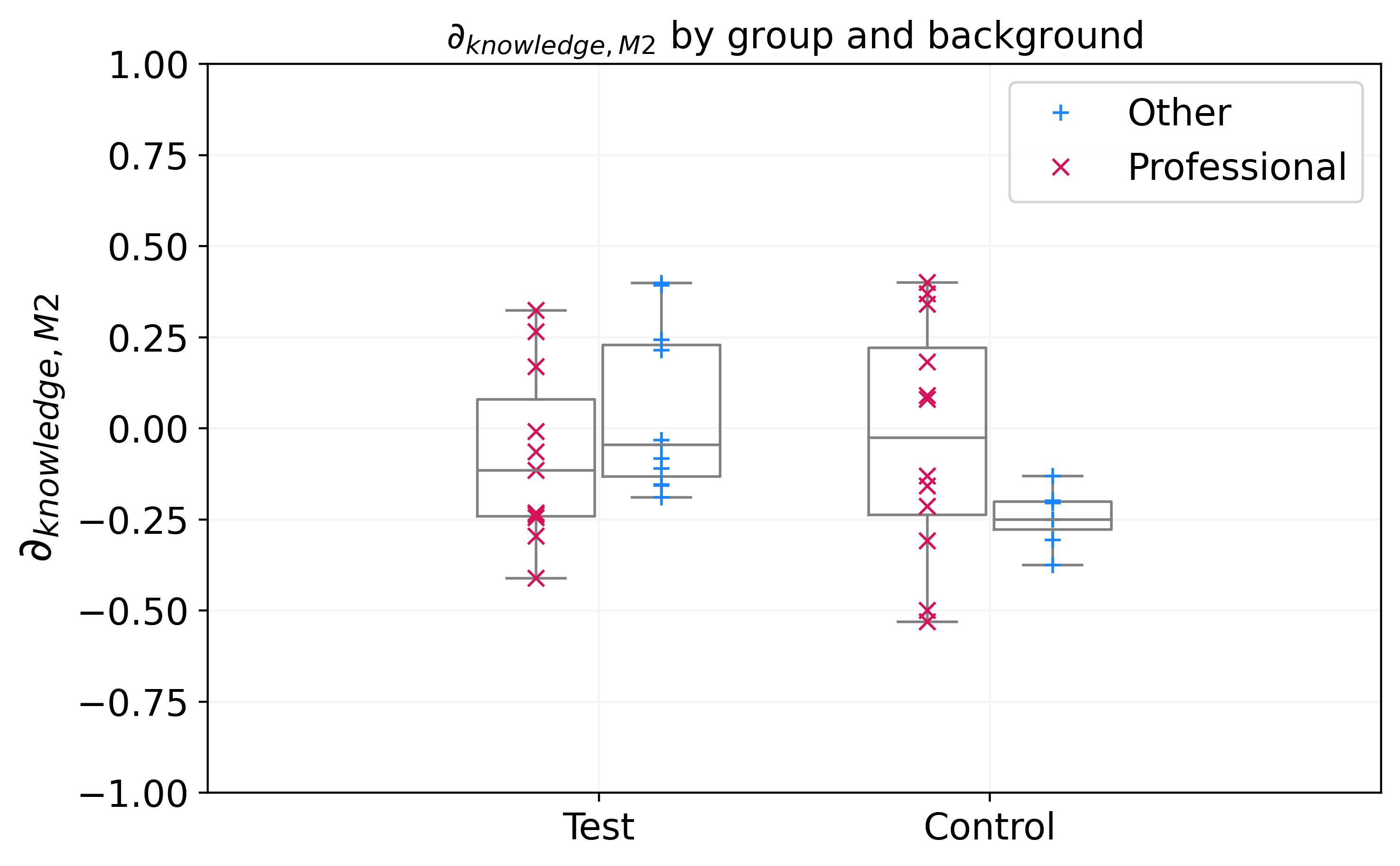} \\
	\caption{Gained knowledge $\partial_{knowledge}$ by group and background (Method 2).}
	\label{fig:Results_Knowledgegain_M2}
\end{figure}

Looking at $\partial_{knowledge}$, we find that the mean $\partial_{knowledge,M1}$ (using method 1, cf. Fig.~\ref{fig:Results_Knowledgegain_M1}) amongst the test group was 0.519 and 0.445 for the control group. The mean of $\partial_{knowledge,M2}$ (using method 2, cf. Fig.~\ref{fig:Results_Knowledgegain_M2}) was 0.018 for the test group and -0.182 for the control group. The distributions look very similar for both methods of calculating $\partial_{knowledge}$. The differences in $\partial_{knowledge}$ (for both methods) between test and control groups are smaller than $\alpha = 0.05$ over the whole sample, cf. Tab.~\ref{tab:StudyResttestVals}. The differences are especially significant for non-professionals, with $p$-values of 0.001 (method 1) and 0.0006 (method 2). For professionals however, there is no significant difference for $\partial_{knowledge}$ between the test and control group with either method.

\subsubsection{Results for Research Question Q2: How does prior experience affect the VR learning gains?}  \label{sec:Results_Study_LearningOutcomes_Q2}
This research question investigates the influences of both prior experiences: professional background and VR familiarity.

The average score achieved by each participant in the exam strictly lies above 0.5 for all participants, cf. Fig.~\ref{fig:Results_AvgScore_Corr} (left). This could be expected because someone filling out the exam totally randomly would still achieve 0.5 points on average. Therefore $s_{exam}$ is calculated according to Eq.~\ref{eq:partialKnowl_m1}, resulting in a broader distribution of the exam scores as shown in Fig.~\ref{fig:Results_AvgScore_Corr} (right). A weak positive correlation ($R = 0.38$) between pre-existing knowledge $s_{PEK}$ and performance in the exam $s_{exam}$ can be observed, cf. Fig.~\ref{fig:Results_AvgScore_Corr}. It also becomes apparent in both graphs, that the distribution of $s_{PEK}$ is very different for the two groups of "Professionals" and "Others". Interestingly, most of the professionals rated their pre-existing knowledge similarly (as measured by $s_{PEK}$) but performed quite differently on the exam (as measured by $s_{exam}$). The test statistics in Tab.~\ref{tab:StudyResttestVals} further show, that $s_{PEK}$ between the two samples is significantly different with a negative test statistics, meaning the PEK was higher in the control group than in the test group. For $s_{exam}$ as well as the other variables listed above, we can see that they are not significantly different between the test and control groups.

Even though the median and average $\partial_{knowledge}$ for professionals do not vary greatly between test and control group, we can observe a smaller distribution for the test group. For non-professionals, a clearly higher mean and median can be observed in the test group, while the distribution is also slightly wider there.

\subsubsection{Results for Research Question Q3: How is the VR tool perceived by students, instructors but also AEC professionals towards ergonomy and usefulness as well as its future potential in education.}  \label{sec:Results_Study_LearningOutcomes_Q3}
This section reports the results towards research question Q3, which deals with the perception of the VR tool and future potentials in education.

\paragraph{Time spent learning}  \label{sec:Results_Study_LearningTimes}
The time spent in the VRE is similar for both groups with a mean of 17.1 (SD=3.51) minutes for the test group and a mean of 16.0 (SD=3.62) minutes for the control group. The time spent studying on paper by the control group however, is much lower with roughly 10 minutes on average.

\begin{figure} [h]
	\centering
	\includegraphics[width=0.98\linewidth]{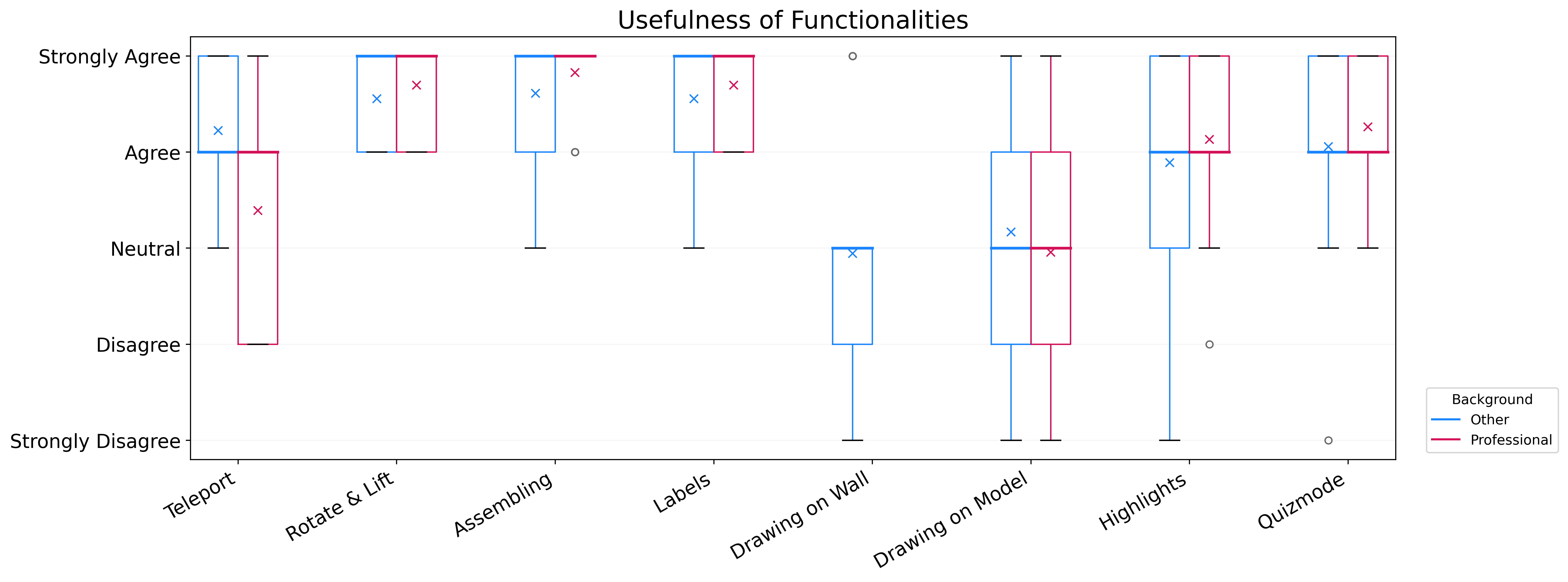} \\
	\caption{Ratings for usefulness of features.}
	\label{fig:Results_FeatureUsefullness}
\end{figure}

\paragraph{Usefulness of Features}  \label{sec:Results_Study_FeatureUsefullness}
In the survey, participants rated the different functionalities of the app according to their usefulness, cf. Fig.~\ref{fig:Results_FeatureUsefullness}. The features rated with highest values are the ones related to the model and the corresponding labels. These features also possess the lowest standard deviations. The lowest ratings were given to the drawings on the wall as well as on the model, yet a high standard deviation can be obtained here. The feature ratings of both, professionals and others, are very similarly in most cases.

\begin{figure} [h]
	\centering
	\includegraphics[width=0.98\linewidth]{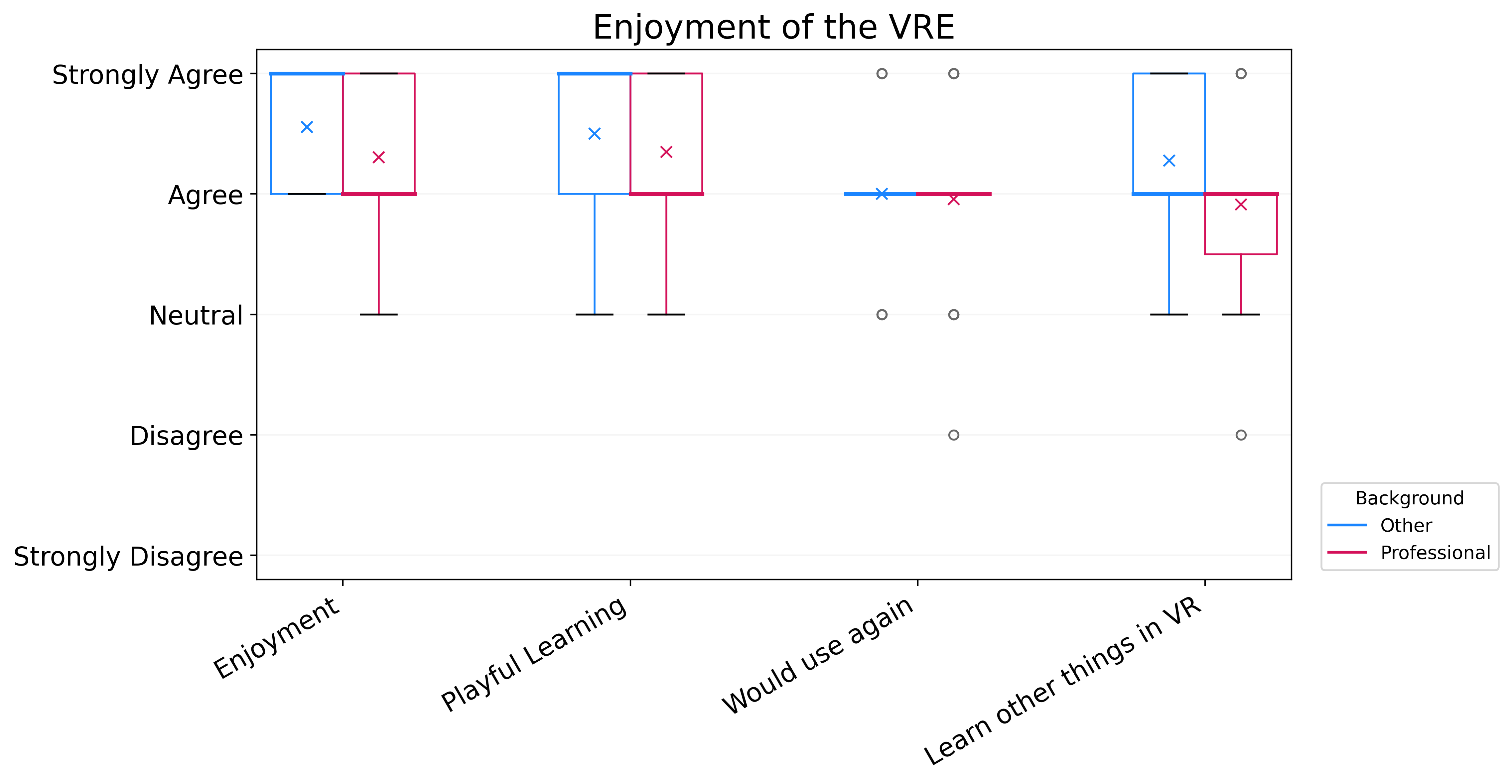} \\
	\caption{Enjoyment ratings for the VRE.}
	\label{fig:Results_VREEnjoyment}
\end{figure}

\paragraph{Enjoyment of the VRE}  \label{sec:Results_Study_VREEnjoyment}
When asked questions about the panellists' enjoyment of the VRE, most users gave high ratings overall, cf. Fig.~\ref{fig:Results_VREEnjoyment}. When asked whether the VRE was enjoyable, the lowest rating given for enjoyment was “neutral” and more than 75\% of users stated “agree” or “strongly agree”. Similar responses were received from the panellist for VR being a playful experience. When asked whether the participants would use VR again to learn, or whether they would like to learn other things with VR, some participants indicated to “disagree”, while most of them still would “agree” or “strongly agree”. The enjoyment ratings of both, professionals and others, are very similarly in half of the cases and differ towards "enjoyment" and "learn other things in VR".

\paragraph{VR in Education}  \label{sec:Results_Study_VREdu}
The participants were asked about their perception of the future of VR in education. The answers were slightly less enthusiastic, but nevertheless very positive. More than 75\% of participants "agree" or "strongly agree" that VR is useful for learning and that it improves learning. While most participants also agreed that VR is useful for teaching complicated concepts and that it is useful alongside a lecture, some strongly disagreed that VR could be used in teaching without an accompanying lecture, especially with a professional background. Participants also agreed less when asked whether VR is better than 2D for learning or whether the 3D model on a computer screen would have been as useful as the VRE. However, most participants agreed that a construction detail can be better displayed in VR than in 2D formats.

\begin{figure} [h]
	\centering
	\includegraphics[width=0.98\linewidth]{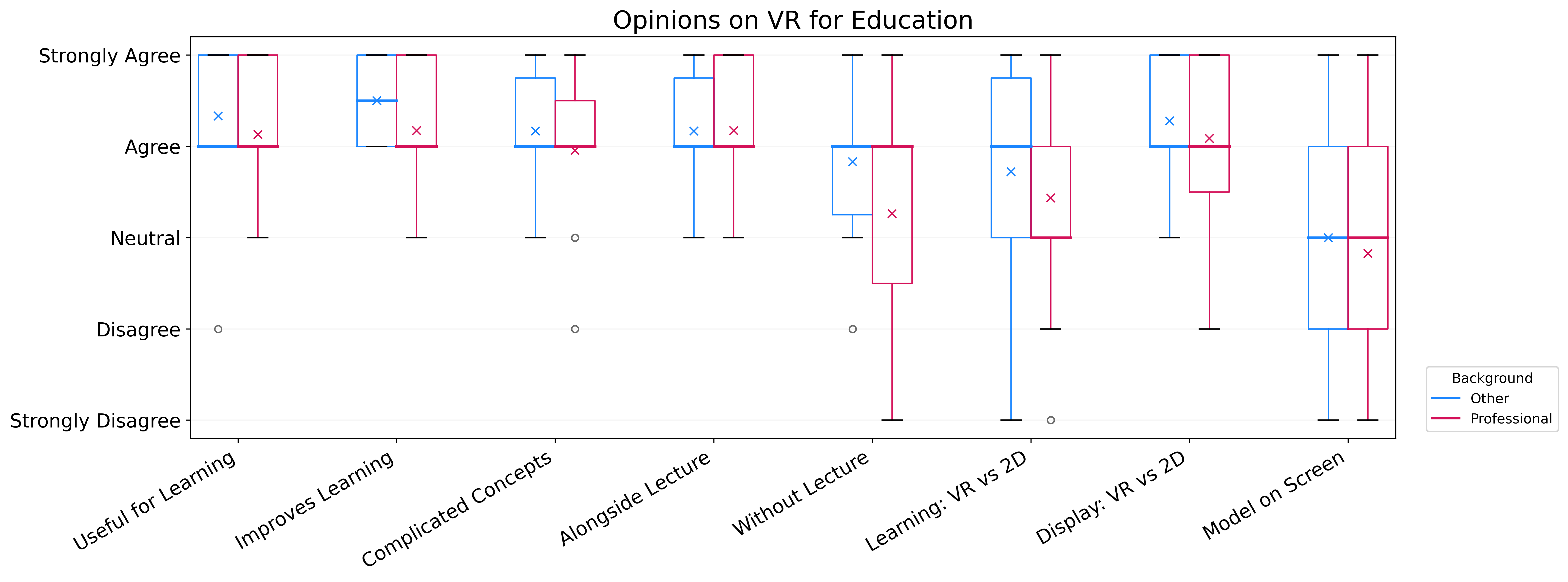} \\
	\caption{Opinions on VR in Education.}
	\label{fig:Results_VREdu}
\end{figure}

Finally we asked the participants in which other areas of AEC education they can imagine potential for VR and its implementation into the curriculum. Most participants chose "construction", then "design studios", which is followed by applications for "structural design" and "building physics". VR was seen least for teaching "architectural history", which today deals mostly with 2D media.

\begin{figure} [h]
	\centering
	\includegraphics[width=0.7\linewidth]{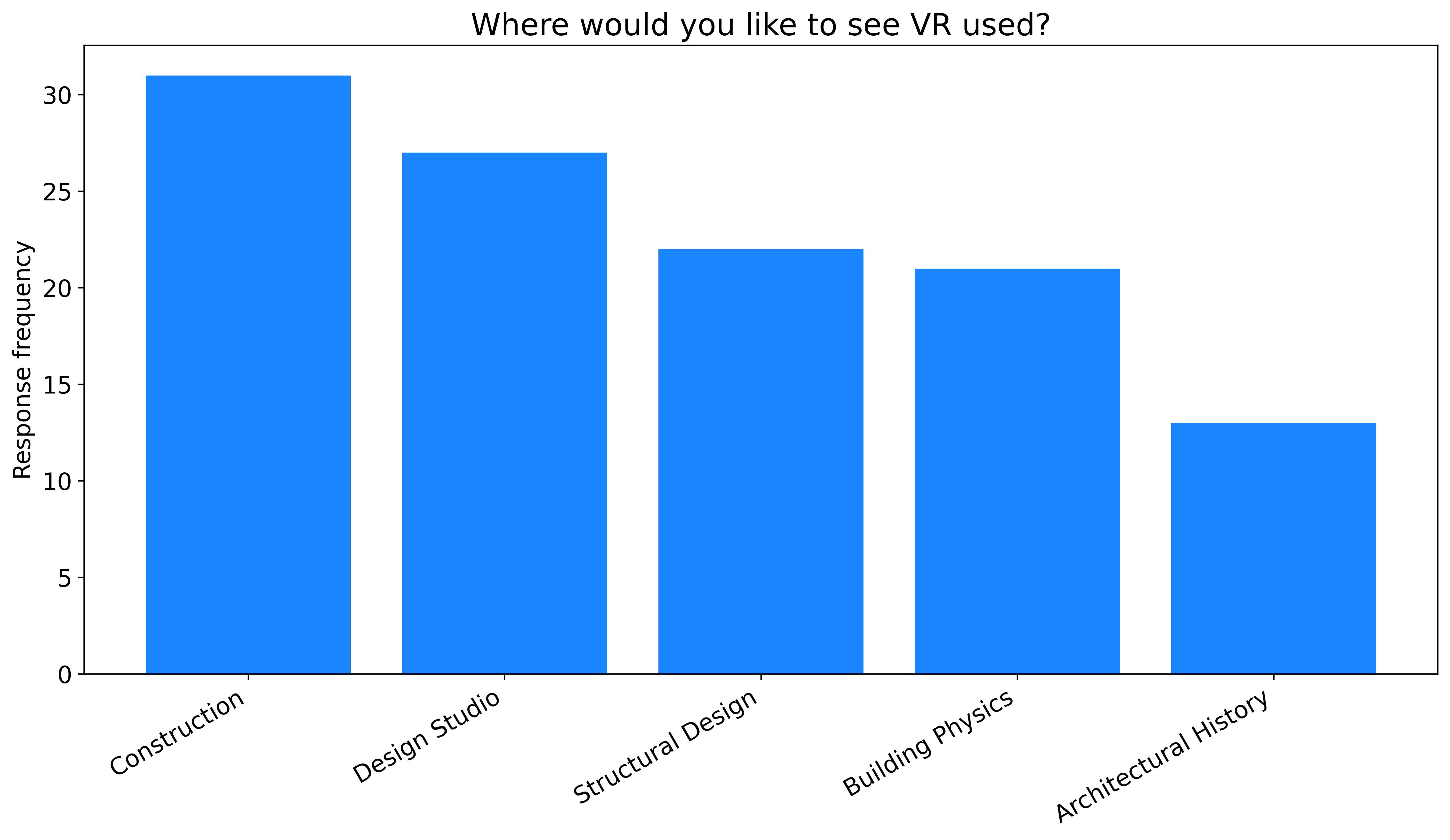} \\
	\caption{Other subjects where participants see VR as useful.}
	\label{fig:Results_VREdu}
\end{figure}

\section{Discussion and Conclusions} \label{sec:discussionConclusion}
This section provides a discussion of the results towards our iterative implementation phase as well as the VRE experiment with associated results.

\subsubsection{Limitations} \label{sec:Limitations}
A limitation of this study is the generalisability of the impact of our VR tool on students' learning as the panellists were not randomly chosen under a uniform distribution across AEC disciplines and beyond, but rather belonged to a selected and small pool of potential participants from the Design++ network at ETH Zurich. While there were more participants in this study than in those reviewed in Sec.~\ref{Background} of this work, this selection bias cannot be quantified further within this paper. On the other hand, convenience sampling in this type of setting is a common procedure, where the random assignment should mitigate bias (which we see to be the case as the p-values are very significant). For establishing a broader database, the repetition of the study in upcoming semesters is planned. In contrast to other studies of XR methods in AEC teaching such as \cite{fogarty2018improving} however, this research established a true control group similarly to \cite{johnson2014collaborative}, which was taught without VR access prior to examination. Due to ethical concerns, the control group was also given access to the VR models after the study in order to avoid any learning shortcomings at their end. Furthermore, we employ triangulation \cite{patton1990qualitative} via different forms of qualitative and quantitative data collection approaches in order to resolve validity concerns for all three study objectives.

\subsubsection{Differences between Methods}
\begin{figure} [h]
	\centering
	\includegraphics[width=0.97\linewidth]{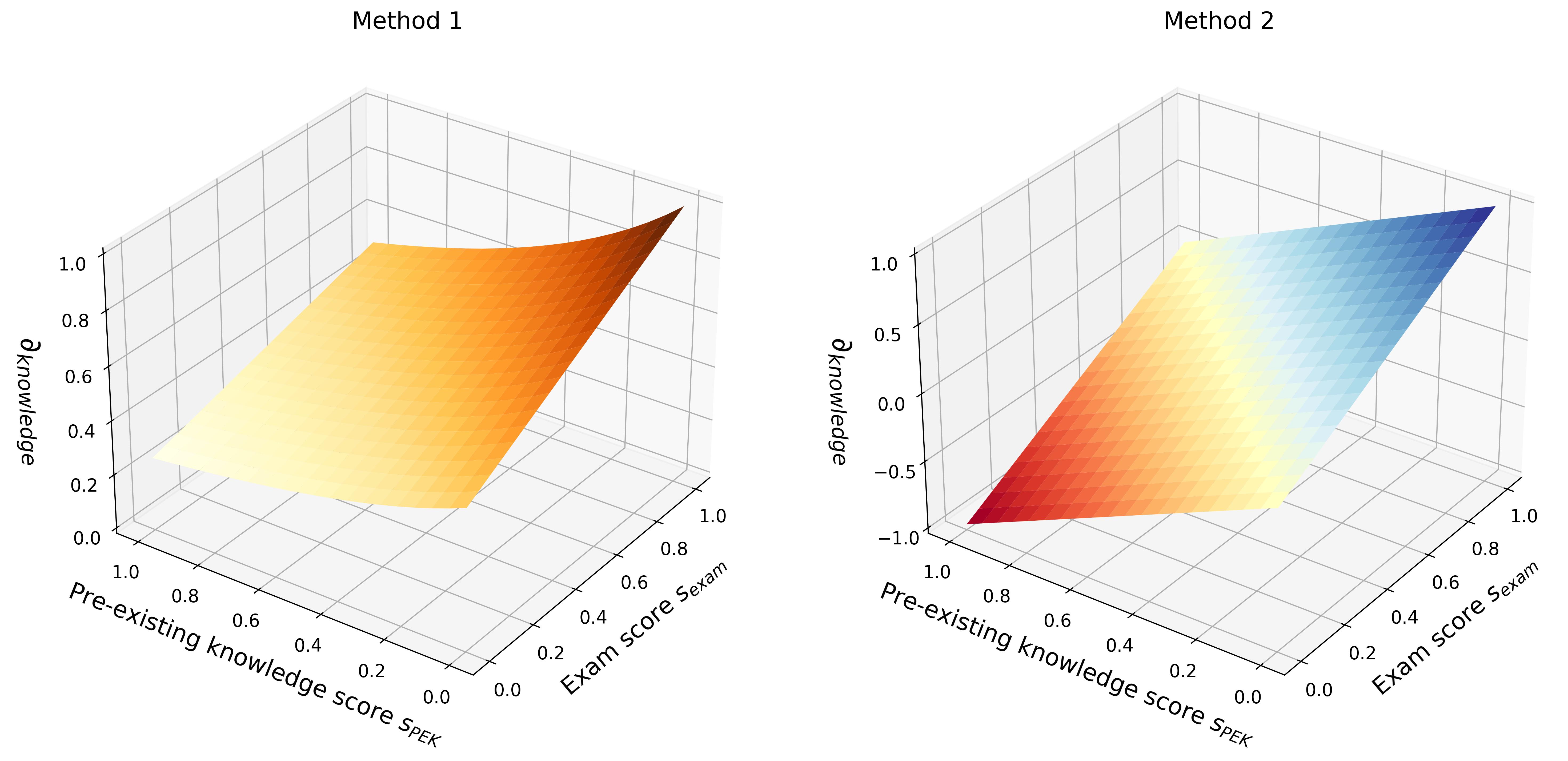} \\
	\caption{3D plots of the knowledge gain $\partial_{knowledge}$ function according to: (left) method 1, and (right) method 2.}
	\label{fig:KnowledgeGainMeasures}
\end{figure}

As can be seen from Tab.~\ref{tab:dknow_Method_1} and Tab.~\ref{tab:dknow_Method_2} and illustrated in graphical form in Fig.~\ref{fig:KnowledgeGainMeasures}, the different knowledge gain measure functions for computing $\partial_{knowledge}$ result in different scales. While the first method maps to an interval between 0.25 and 1 with the point for ‘no learning observed’ at 0.5. The range for a positive learning outcomes is double as large as the range for no learning. Therefore improved learning has a larger deviation from the ‘no learning point’ than 'no learning'. The second method gives equal range to the two situations as this function is linear. A value for $\partial_{knowledge}$ that is lower than 0.5 with method 1 or lower than 0 with method 2 means that the participant performed worse than could be expected with their pre-existing knowledge. Since it is unrealistic to assume that participants un-learned something during the trial, a large number of such values would indicate that $s_{PEK}$ is inaccurate and that the questions for calculating $s_{PEK}$ should be adjusted.

\subsubsection{Discussion and Conclusions for Research Question Q1: Does VR improve student's understanding of complex spatial construction arrangements in contrast to 2D drawings?}  \label{sec:Discussion_Q1}

Within the discussion, we expect students to be non-experts (hence we associated them with the "Other" group during the computation of the results), VR significantly improves student's understanding. 
The results for the two calculation approaches for $\partial_{knowledge}$ are very similar and hence lead to similar outcomes in terms of the hypothesis tests. To verify whether the test groups higher values for $\partial_{knowledge}$ are statistically significant, they were tested in a two sample t-test. From the results of these tests we can conclude that the difference in learning effect is statistically significant (and hence research question Q1 holds true), except when looking at the professional group only. The $p$ values across all participants are 0.0209 and 0.0328 for methods 1 and 2 of calculating $\partial_{knowledge}$ respectively. Therefore the null hypothesis can be rejected and we can say the difference in learning outcomes is significant. The same is true for the $p$ values of 0.0010 (method 1) and 0.0006 (method 2) in the group "Others". However, for professionals, the $p$ values are 0.9161 and 0.775 and therefore the null hypothesis can not be rejected in that case, so we have to assume that the VRE did not improve learning for them. However, the fact that for professionals the scores are more widely distributed (cf. Fig.~\ref{fig:Results_Knowledgegain_M1} and Fig.~\ref{fig:Results_Knowledgegain_M2}) in the control group indicates that even though on average they have not learned more with the VR, their performance has become more streamlined. We can deduce that preexisting knowledge must have played less of a role in the test group than in the control group. We can conclude that, while it might not be significant, a larger learning effect has also occurred for professionals in the test group compared to the control group. From that, we can conclude, that research question Q2 can be positively answered. 

With the results from the survey it becomes clear that the participants preferred the VRE over learning the traditional way. The fact that most participants in the control group spent more time in the VRE than for learning on paper, even though they had already completed the exam, is indicative of the attractiveness of the VRE. The VRE, with its playful elements, can keep the users attention focused on the matter much longer than a drawing on paper. It would have been interesting to add a third group of participants which is allowed to explore the same model as the VR group, not in a VRE but on a computer screen. Standard computers share some of the advantages with VR technology over 2D drawings w.r.t. graphical illustration of spacial objects. Yet computer visualisations still are missing the elements of immersion and part of the interactivity, but may still benefit students learning of construction details. Interestingly this idea was on average rated neutral or slightly disagreed with by the participants with a relatively high standard deviation. This could be related to participants' experiences: some had difficulties in navigating the VR experience and then might have vome to the impression, that VRE has no advantage over a computer screen while other panellists felt comfortable in the VRE and hence rated VR the better medium.

This VRE came without a human instructor but provided a tutorial. Earlier studies of \cite{fogarty2018improving,birchfield2009earth} stress the importance of face-to-face interaction and discussion of the concepts during the VR experience. Future developments of this VRE will consider scenarios along these lines. Adopting a slightly broader view, VR technology in general and VRE such the one of this study in particular could not only help students in the fields of AEC but also laypeople (e.g. clients) to understand architectural or structural configurations much faster, better, and more intuitive. Finding effective means of improving also the learning with professionals through VRE is subject to future studies.

\subsubsection{Discussion and Conclusions for Research Question Q2: How does prior experience affect the VR learning gains?}  \label{sec:Discussion_Q2}

The correlation between $s_{PEK}$ and $s_{exam}$ is pronounced ($r = 0.38$), cf. Fig.~\ref{fig:Results_AvgScore_Corr}, which implies that the pre-existing knowledge cannot be ignored as influencing factor for the learning outcome measured via $s_{exam}$. The results of the tested hypotheses as provided in Sec.~\ref{sec:Results_Study_LearningOutcomes_Q2} underline this, cf. Tab.~\ref{tab:StudyResMeanVals} as well as Fig.~\ref{fig:Results_AvgScore_Corr}. On the other hand, the eminent variance within the "professional" group for the $s_{PEK}$ score indicates, that the current approach to quantify the prior knowledge is imprecise to some degree. The subjective rating nature of the present survey could be improved towards more objectively assessing $s_{PEK}$ by adjusting the questions in the survey for pre-existing knowledge or by providing a pre-quiz with similar questions to the exam. Another approach is to modifying the formula for computing $s_{PEK}$. One core assumption of computing $s_{exam}$ was, that none of the panellist intentionally cheated when providing answers during the exam, yet filling the true/fals questions on average would lead to a 50\% score. In order to reduce further the influence of this, more questions (especially of the multiple-choice or open ended type) can be added to the exam.

The VRE is better for learning when participants have limited or no pre-existing knowledge at all. This is the case for laypeople but more importantly for freshman AEC students, hence the VRE is a suitable medium in early phases of AEC curricula. The fact, that people with professional backgrounds did not benefit significantly from the VRE, is interesting. It could stem from the fact that there was just one construction detail shown in the VRE. Furthermore, as the construction detail example was taken from an existing course, some professionals may already have been familiar with it and hence bias the outcomes. The study could be repeated with more complex or otherwise new details, that contain information that most participants do not know a priori. In that case, one could expect the results to be similar to the results for the group "Others" in this study, as the circumstance that the participant has never seen such a detail before would then also be true for participants with a professional background.

\subsubsection{Discussion and Conclusions for Research Question Q3: How is the VR tool perceived by students, instructors but also AEC professionals towards ergonomy and usefulness as well as its future potential in education.}  \label{sec:Discussion_Q3}

The received general feedback from participants during the study is in line with these findings, where the ability to interact with the model (features "Rotate \& Lift" and "Assembling" in Fig.~\ref{fig:Results_FeatureUsefullness}) and the consultancy of the tablet resp. 3D model information features ("Labels" in Fig.~\ref{fig:Results_FeatureUsefullness}) were highlighted as the main benefits. The functionalities that enable interaction with the model are rated highest by the users. Therefore we can conclude that the model itself and the interactions with it are the most useful part of the whole VRE. At this point we further conclude, that embodied learning in the VR environment plays a role in the perception and understanding of the teaching material. Several panellists remarked on the advantage of being able to see the dynamic process of assembling the construction detail. This is in line with the findings of \cite{fogarty2018improving}, where students also saw a positive effect through dynamic animations (which is in contrast to the comment in \cite{tversky2002animation} about animations tend to not actually be useful when compared with a static version of the same phenomenon even if it is conveying a change over time). The majority of participants agreed or strongly agreed with the respective statements labelling the VRE as enjoyable and playful, where not a single respondent disagreed with these statements. More than 75\% of participants also agreed or strongly agreed that they would use the VRE for learning again or that they would like to learn other things with VR. These numbers clearly show what could be observed qualitatively in the study as well - people enjoy trying VR experiences.

Removing all elements around the model (drawings, tutorial, quiz mode, tablet) could result in a similarly high learning effect, if the controls are intuitive enough. Therefore it is worthwhile to discuss removing certain functionalities, starting with the ones rated least useful such as the 2D drawings or the object highlights. If such a simplification would allow to reduce the complexity of the interface, for example by not needing a tablet, this could help the users to better focus on the model. Users that have used VR before would likely navigate such a reduced version of the app easily, while users that have never used VR before might struggle more with such an app, even though they would also benefit from the removal of distractions. On the other hand a standardised, well designed user interface would make sense in an application that users use more than once. If they have time to get used to the way a VRE works and to familiarise themselves with the interface, a well designed interface would definitely allow for more functionalities to be implemented without risking overwhelming the users or distracting them from the subject matter. When developing such a VRE that covers many construction details, it would be worth investing time into such a well designed interface to leverage some extra functionalities to improve the learning effect.

VR applications are still not common amongst significant parts of society, where half of our participants stated to have never used a VR device before. This definitely works to the benefit of the above mentioned metrics, however, the effect can surely not be ascribed solely to that reason. Since VREs are entertaining and remind users of games, they are inherently more fun and engaging than traditional learning methods. It is therefore also understandable that the majority of users would use the VRE again or would like to learn other things with VR. In the future, when VR technology becomes more common in our daily lives and in education, these metrics might change slightly, but it is unlikely that users will start disliking VREs as a means of learning. The user acceptance of VR technology in teaching can therefore said to be high. In general we observed that most participants see clear advantages in learning with VR. However, from some questions' responses we can see that users are not blindly in love with VR. Not a single participant disagreed with the notion that the VRE would be useful as a supplement to a lecture series and on average participants agreed with this idea. When asked whether the VRE would be useful as the main means of teaching, without an accompanying lecture series, some strongly disagreed and on average the answer was less enthusiastic somewhere between neutral and agreement. The participants on average agreed that VR is a better means than 2D drawings for showing construction details. This is not a surprise, since as mentioned in the introduction, 2D drawings are always abstractions of a 3D context that contain less information than a 3D model of the same configuration. Surprisingly the participants had a different view when asked whether VR was better for learning than 2D drawings. The average response was between neutral and agreement here. When considering the fact that professionals did not learn better in the VRE than on paper, this makes some sense. Another factor could be that professionals are already used to learning with drawings but not familiar with VR. Regardless of the learning outcome, the users might prefer VR technology for learning, simply because it is more entertaining and engaging. This effect can be seen amongst both groups, professionals and others.

Looking at the participants opinions about the VRE and VR in education, it can be said that the user acceptance for VR within the educational sector is quite high. Combined with technological advancement and even decreasing costs for implementing VR technology, this will lead to the introduction of VR into more and more areas within education in general and AEC education in particular. Furhtermore, XR integration into AEC curricula prepares students better for the demands of the 21st century industry and the reported approach in this paper specifically enables them to exceed boundaries of traditional learning through immersive learning experiences and fostering the interest in the XR technology, which then is conveyed into practice through the educated students and instructors.

The survey results indicate directions for further development of a VRE like the one presented in this work. Simplicity is key, especially for the foreseeable future, where the majority of users will be inexperienced with VR. Intuitive interaction is also of foremost importance to leverage the technology and to exploit its possibilities in the best possible way. The VRE discussed here was a simple prototype with limited scope. There are many functionalities that could be added to future versions, such as letting the user construct the detail themselves by manually placing building elements, by including physics for understanding the structural design or by implementing visualisations to explain thermal bridges for example. We can conclude that VR definitely has a place in future AEC curricula but it is certainly not a solution that will overtake AEC education completely. Especially in the near future, it will enjoy popularity as a supplemental means of teaching in combination with other teaching formats. Nevertheless, the study revealed convincing arguments for implementing VR technology into architecture education. 3D models carry more information in a way that is easier to understand than traditional teaching materials. In addition, VR experiences enable new ways of learning, problem-based learning and experiential learning in particular, which are definitely superior to the traditional way of teaching, which relies heavily on memorising facts. Interaction and engagement are important factors for learning success and these are the areas where VR technology shines.

\section{Outlook}  \label{sec:Conclusion}
%Despite all negative aspects, the COVID 19 pandemic 
With the recently increasing awareness for digital teaching methods and formats and their experimental introduction into curricula (such as at ETH Zurich since the past semester), we hope to see other higher educational institutions following this approach to increase experience and references. %. Yet the lockdowns during the pandemic trajectory also imposed a sharp switch to these new means of education without a phase of prior research and gathering of experience. 
The mixed-methods study of this paper looks at the usage of VR technologies in the classroom and how they might help students grasp complex spatial arrangements with the example of a basement detail. Based on the authoring teams' knowledge in instructing students, the VRE together with the exam and survey was devised. Significance tests revealed that students increased their test scores and consequently their understanding of the construction detail on average. The ability to physically interact with the model within the VRE and the use of additional information features were the main aspects of the models that contributed to improving student understanding of the given spatial configurations, according to this study.

To see if the obtained results are actually generalisable, all of these elements should in the future be transferred to other construction details from AEC but also other universities. As technology advances and VR technologies become more inexpensive and widely available, current and future models can be shared with educators globally and applied in a wide range of courses using simple scripting to convert models into usable data for the VR environment. Further research into the effects of immersion and embodied learning will also be carried out. Additional virtual reality tools will be developed and tested to see if the characteristics of these VR models that have been found as important in aiding student learning are generalisable to other complicated spatial concepts.

\section{Appendix}  \label{sec:Appendix}
This appendix compiles some additional information for the control group in the following table.

\begin{table}
	\caption{Some additional information for the control group}
	\label{tab:Annex2}
	\centering
	\small
	\renewcommand{\arraystretch}{1.25}
	\begin{tabular}{l l}
		\hline\hline
		\multicolumn{1}{c}{Element} & \multicolumn{1}{c}{Explanation} \\
		\hline
		Terrain & To construct the basement an excavation is prepared. \\
		Drainage Pipe       & The sloping drainage pipe removes water \\ & that might be accumulating here. \\
		Gravel              & Gravel lets any water run down quickly.                                  \\
		Geotextile Mat      & A getotextile mat stabilises the ground. It is water-permeable.          \\
		Earth               & Once the Basement has been constructed \\ & the terrain is filled in again.   \\ 
		Concrete Slab       & The concrete slab divides the drainage layer from the garden.            \\ 
		Geotextile Mat      & A getotextile mat stabilises the ground. It is water-permeable.          \\ 
		Coarse Gravel       & Gravel lets any water run down quickly.                                  \\ 
		Humus Layer         & The humus layer allows plants to grow.                                   \\ 
		Grass               & -                                                                        \\ 
		Lean Concrete       & The porous lean concrete is for preparation \\ & and to level the ground.     \\ 
		Concrete Foundation & The concrete plate is poured directly \\ & onto the lean concrete.            \\ 
		Moisture Barrier    & The waterproofing prevents humidity from \\ & entering the building from below. \\
		Insulation (Wall)   & The vapor-tight insulation prevents condensation \\ &  within the wall construction \\
		Masonry             & The Masonry carries the weights of the ceiling above.                    \\
		Insulation (Floor)  & Towards the ground the floor is insulated with a \\ & layer of insulation that is not damaged by water. \\ 
		Separating Foil     & A separating foil is laid onto the insulation \\ & to prevent the UB from leaking below. \\ 
		Concrete Screed     & The concrete screed can contain underfloor heating.                      \\ 
		Plastering          & -                                                                        \\ 
		Floor Finish        & -                                                                        \\ 
		Concrete Wall       & -                                                                        \\ 
		Exterior Plastering & The plastering here is waterproof and prevents \\ & humidity from entering the construction.        \\ 
		Drainage Panels     & Drainage Panels are porous and ensure that water can run off quickly.    \\
		\hline\hline
	\end{tabular}
	\normalsize
\end{table}

\newpage
\section{Data Availability Statement}
% All data, models, or code generated or used during the study are available in a repository online in accordance with data retention policies: \href{https://github.com/rbischof/relative_balancing}{https://github.com/rbischof/relative\_balancing}

Some or all data, models, or code that support the findings of this study are available from the corresponding author upon reasonable request (list items) and will be made publicly available after publication here: \href{https://github.com/mkrausAi/VR_Teaching_Construction}{https://github.com/mkrausAi/VR\_Teaching\_Construction}.

\section{Acknowledgment}
The authors would like to acknowledge the campus facilities of the Design++ Initiative of ETH Zurich and especially Profs. Walter Kaufmann, Fabio Gramazio and Matthias Kohler for providing financial support, and devices for the developments presented in this paper. Furthermore the discussions and contributions of M.Sc. Irfan \v{C}ustovi\'{c} are highly appreciated.

\section{Notation list} \label{app:notation}
\emph{The following symbols are used in this paper:} 
\nopagebreak
\par
\begin{tabular}{r  @{\hspace{1em}=\hspace{1em}}  l}
	$n_{ex,correct}$    & number of correctly answered exam questions (-); \\
	$n_{ex,total}$      & total number of exam questions (-); \\
	$s_{exam}$          & exam score (-); \\
	$<s_{exam}>$        & average exam score (-); \\
	$s_{PEK}$           & pre-existing knowledge score (-); \\
	$\partial_{knowledge,i}$ & knowledge gain as relationship between $s_{exam}$ and $s_{PEK}$ \\ & computed via method $i$ (-); \\
	VR Quiz Score    & average score of points obtained in the quiz mode for \\ &  panellists of the control group (-); \\ 
	VR PEK Score     & score for VR familiarity (-); \\ 
	VR UX Score      & score for user experience within our VRE (-); \\ 
	VR Liking Score  & score for user enjoyment of our VRE (-); \\ 
	VR in EDU Score  & score for user perception of VR in education (-); \\ 
\end{tabular}

\section{Supplemental Materials}  \label{sec:Supplemental}
The supplementary material to this research consists of the list of questions of the conducted survey as well as the list of examination questions. The files can also be obtained via \href{https://bit.ly/3w8gXyd}{https://bit.ly/3w8gXyd}

\bibliographystyle{acm}
\bibliography{ASCE_KrausEtAl}

\end{document}